\documentclass{article}
\usepackage{natbib}
\usepackage{fullpage}
\usepackage{graphicx}
\usepackage{color}
\usepackage{xcolor}
\definecolor{noPeaks}{HTML}{F6F4BF}
\definecolor{peakStart}{HTML}{FFAFAF}
\definecolor{peakEnd}{HTML}{FF4C4C}
\definecolor{peaks}{HTML}{A445EE}
\newcommand{\annsquare}[1]{
  \textcolor{#1}{\rule{0.1in}{0.1in}}
}
\usepackage{amsmath,amssymb}
\DeclareMathOperator*{\argmin}{arg\,min}

\DeclareMathOperator*{\minimize}{minimize}
\newcommand{\RR}{\mathbb R}

\newcommand{\figtrain}{Supplementary Figure S1}
\newcommand{\figroc}{Supplementary Figure S2}
\newcommand{\figlearned}{Supplementary Figure S3}
\newcommand{\figseveral}{Supplementary Figure S4}
\newcommand{\figannotators}{Supplementary Figure S5}
\newcommand{\figtypes}{Supplementary Figure S6}

\usepackage{hyperref}

\begin{document}

\title{Visual annotations and a
  supervised learning approach for evaluating and calibrating ChIP-seq peak
  detectors}

\author{%
Toby Dylan Hocking (toby.hocking@mail.mcgill.ca), \\
Patricia Goerner-Potvin,
Andreanne Morin,
Xiaojian Shao,
and Guillaume Bourque
\\
  Department of Human Genetics, McGill University, Montr\'eal, Canada.\\
  McGill University and G\'enome Qu\'ebec Innovation Center,
  Montr\'eal, Canada.  }

\maketitle

\begin{abstract}
  Many peak detection algorithms have been proposed for ChIP-seq data
  analysis, but it is not obvious which method and what parameters are
  optimal for any given data set. In contrast, peaks can easily be
  located by visual inspection of profile data on a genome browser. We
  thus propose a supervised machine learning approach to ChIP-seq data
  analysis, using annotated regions that encode an expert's
  qualitative judgments about which regions contain or do not contain
  peaks. The main idea is to manually annotate a small subset of the
  genome, and then learn a model that makes consistent predictions on
  the rest of the genome. We show how our method can be used to
  quantitatively calibrate and benchmark the performance of peak
  detection algorithms on specific data sets. We compare several peak
  detectors on 7 annotated region data sets, consisting of 2 histone
  marks, 4 expert annotators, and several different cell types. In
  these data the macs algorithm was best for a narrow peak histone
  profile (H3K4me3) while the hmcan.broad algorithm was best for a
  broad histone profile (H3K36me3). Our benchmark annotated region
  data sets can be downloaded from a public website, and there is an R
  package for computing the annotation error on GitHub.
\end{abstract}

\newpage

\tableofcontents

\newpage

\section{Introduction and related work}

Chromatin immunoprecipitation sequencing (ChIP-seq) is a genome-wide
assay to profile histone modifications and transcription factor
binding sites \citep{chip-seq}, with many experimental and
computational steps \citep{practical}. In this paper we propose a new
method for the peak calling step. The goal of peak calling is to
filter out background noise and accurately identify the locations of
peaks in the genome.

There are two main lines of research into software tools that can help
scientists find peaks in the genome. One class of software consists of
peak detection algorithms, which are non-interactive command line
programs that can be systematically run on all samples in a data
set. An algorithm takes the aligned sequences as input, and returns
precise locations of predicted peaks as output. Despite these
advantages, peak detector software has one major drawback: model
selection. There are many different algorithms that are specifically
designed for peak detection in ChIP-seq data, each with many
parameters to tune in each algorithm. Each algorithm and parameter
combination will return a different set of predicted peaks. Given a
specific ChIP-seq data set to analyze, how do you choose the best peak
detection algorithm and its parameters?

The second class of software consists of graphical tools such as the
UCSC genome browser \citep{ucsc}. To view ChIP-seq data on the UCSC
genome browser, the ChIP-seq coverage must be saved to a bigWig file
\citep{bigwig}, which can be browsed as a line or bar plot to visually
identify peaks. The main advantage of this approach to peak detection
is that it is often easy to visually identify peaks and background
noise in coverage plots of several ChIP-seq samples. There are two
main disadvantages of this approach. First, precise peak start and end
locations are not obvious on visual inspection. Second, no researcher
has enough time to visually inspect and identify peaks across the
whole genome.

\begin{figure*}[b!]
\begin{center}
\includegraphics[width=\textwidth]{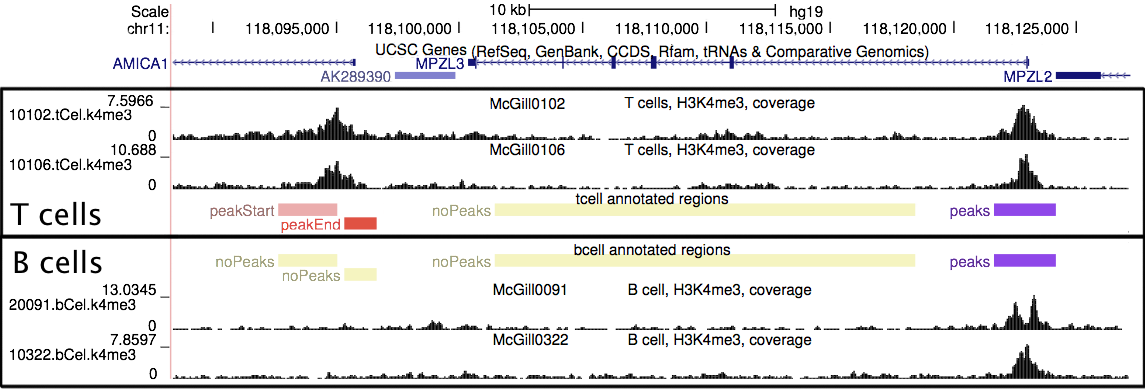}
\end{center}
\vskip -0.5cm
\caption{One genomic window containing 4 annotated regions for each of the 4
  shown profiles. 
  Visual inspection of ChIP-seq normalized coverage plots can be used 
  to create annotated regions that encode where peaks should and should not be
  detected in these T cell and B cell samples. Exactly 1 peak start/end
  should be detected in each peakStart/peakEnd region. There should be
  no overlapping peaks in each noPeaks region, and at least 1 overlapping peak
  in each peaks region. 
}
\label{fig:annotations}
\end{figure*}

In this article we propose a new machine learning approach for
ChIP-seq data analysis that combines these two lines of research. The
main idea is to manually annotate peaks in a small subset of the
genome, and use those annotations to learn a peak detection model that
makes consistent predictions on the rest of the genome. In particular,
we propose to create annotated regions that encode an experienced
scientist's judgment about which regions contain or do not contain
peaks (Figure~\ref{fig:annotations}). The annotated regions can then
be used as a gold standard to calibrate model parameters and then
evaluate peak detection algorithms on specific data sets.


\enlargethispage{-65.1pt}
\subsection{Related work: benchmarking peak detectors}

There are several algorithms for detecting peaks in ChIP-seq data
\citep{MACS, homer, HMCan, RSEG, SICER}, and in this article we
propose to benchmark their peak detection accuracy on specific data
sets using manually annotated regions.  Other methods for benchmarking
ChIP-seq peak callers include using known binding sites
\citep{known-binding-sites}, low-throughput experiments
\citep{low-throughput}, and simulation studies
\citep{chip-seq-bench}. Each of these benchmarking methods has its own
strengths and weaknesses. For example, known binding sites are useful
positive controls for transcription factor ChIP-seq, but are rarely
known for histone marks. An unlimited amount of data can be generated
using computational simulation studies, but these data may be
arbitrarily different from real data sets of interest. Low-throughput
experiments are always useful to confirm binding sites at specific
genomic locations, but are not routinely done to accompany genome-wide
ChIP-seq experiments.

The manual annotation method that we propose in this paper is much
more widely applicable for benchmarking ChIP-seq peak detectors. The
main weakness of our proposed method would be the time required for
manual visual annotation of some genomic regions of the ChIP-seq
data. However, we show in the results section that our method is
useful even if there is only time to create a small annotated region
database.

\subsection{Related work: supervised, interactive analysis}

Supervised machine learning methods have been applied to ChIP-seq
without using interactive data visualization. For example,
low-throughput experiments were used to train several supervised
classification models to improve ChIP-seq peak calling
\citep{low-throughput}. As another example, a supervised machine
learning approach was used to define a regulatory vocabulary with
genome-wide predictive power \citep{chip-seq-ml}.

Other recent software tools focus on interactive visualization of
ChIP-seq data, without using annotations and supervised machine
learning approaches \citep{spark, chase}. They take several profiles
and several genomic regions as input and allow the user to
interactively adjust clustering model parameters. Both are similar to
the approach used for Fluorescence-Activated Cell Sorting (FACS) data
analysis, where the user manually specifies fluorescence thresholds
for sorting cells. 

Our interactive, visual approach to ChIP-seq data analysis is closely
related to several other recently proposed software tools for
supervised machine learning of biological data. For example,
CellProfiler Analyst is an interactive system for semi-automatically
labeling cell phenotypes in high-content cell microscopy screening
assays \citep{cellprofiler}.

To apply these visual methods to genomics data, we earlier proposed to
benchmark breakpoint detection algorithms for DNA copy number analysis
using a database of annotated regions that contain or do not contain
breakpoints \citep{HOCKING-breakpoints}. We further proposed to use
these annotated regions in SegAnnDB, a web site for supervised,
interactive DNA copy number analysis \citep{HOCKING-SegAnnDB}. In the
present paper we adapt this line of research for peak detection in
ChIP-seq data.

\section{Materials and Methods}

\subsection{Annotating samples}

\begin{figure*}[b!]
\begin{center}
  \includegraphics[width=\textwidth]{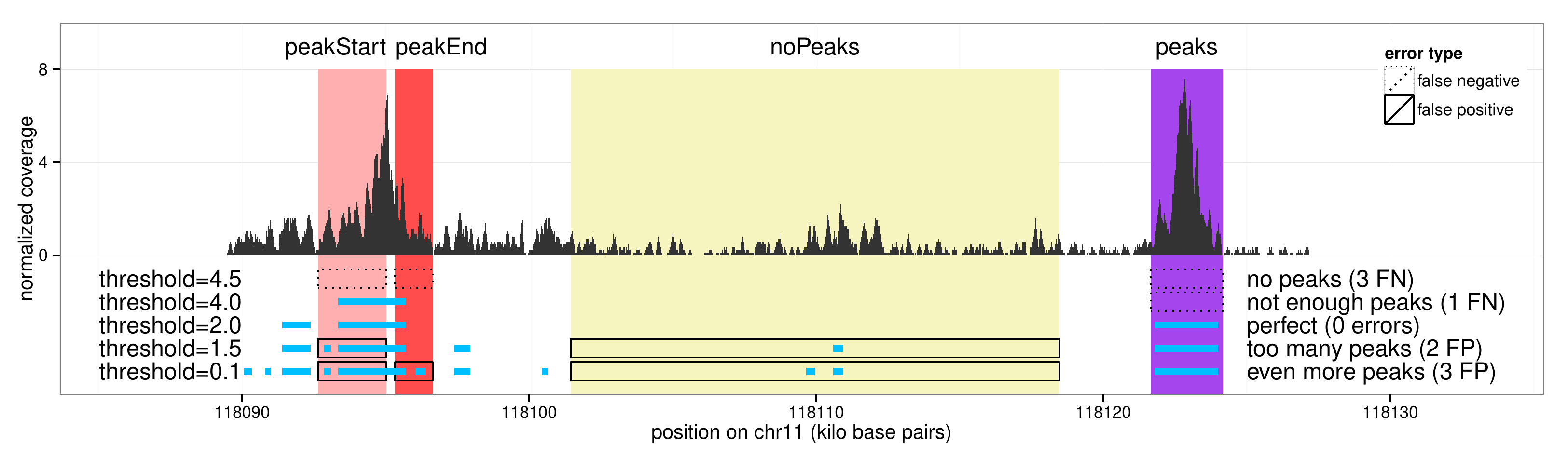}
\end{center}
\vskip -0.7cm
\caption{Annotated regions can be used to quantify the accuracy of a
  peak detection model. Peaks detected by five different thresholds of
  the HMCan model are shown in blue for tcell sample McGill0102, mark
  H3K4me3, annotated by TDH. Models with too few peaks have false
  negatives (threshold$\geq 4$), and models with too many peaks have
  false positives (threshold$\leq 1.5$), so for these data we choose
  an intermediate threshold=2 that minimizes the number of incorrect
  annotations.}
\label{fig:PeakError}
\end{figure*}

The first step of analysis is to create a database of annotated
regions, by visually inspecting coverage plots of the ChIP-seq
samples. It is advantageous to simultaneously inspect several samples
of a single experiment type, to more easily identify common peaks and
noise across multiple samples (Figure~\ref{fig:annotations}). It is
useful to also simultaneously view supplementary data tracks such as
genes, alignability/mappability, and related input/control samples. To
ensure the creation of a gold standard annotated region database of
high quality, we only annotated regions which were very obvious, and
avoided annotating any regions which were unclear. The boundaries of
the regions can be made as large or small as necessary.

For each annotated region, we copied the genomic coordinates to a text
file and noted the annotation (example file in supplementary
materials). As shown in Table~\ref{tab:colors}, we considered four
types of annotations: peakStart, peakEnd, peaks, and noPeaks.  Each
noPeaks annotation is used to designate a region that definitely
contains only background noise, and contains no peaks. When a peak
start or end is visible, it can be annotated using a peakStart or
peakEnd region. These regions should contain exactly one peak start or
end, and do not necessarily need to occur in pairs. For example, if
the peak start is clear and the peak end is unclear, then one should
add a peakStart region, and not add any nearby peakEnd
region. Finally, a peaks region means that there is at least one
overlapping peak. These peaks regions are useful when the number of
peaks, or the start and end locations are unclear. For example, on the
right side of Figure~\ref{fig:annotations} it is clear that there is
at least one peak, but in profile McGill0091 there seems to be two
peaks. So we created a peaks region, which means that either one or
two peaks in that region is acceptable (but zero peaks is
unacceptable). 

We observed similar peaks across samples of a given cell type, and
always assigned the same annotated regions to each of those samples
(Figure~\ref{fig:annotations}). So that we could later review and
verify the annotated regions, we grouped our annotation database into
windows of nearby regions, such as the window shown in
Figure~\ref{fig:annotations}. We made sure that each window contains
at least one region with a peak and one region without
peaks. Furthermore, we made sure that no region overlaps any other
region on the same sample.

Finally, because each window contains the same annotated regions for
samples of the same cell type, it is appropriate to use windows rather
than regions as units of cross-validation. So in the computational
experiments where we measure the peak detection test error, we train
on several windows of annotated regions, and test on several other
windows.





\subsection{Annotation error function and peak detection problem}

Assume we have $n$ annotated training samples, all of the same
ChIP-seq experiment type. For simplicity, and without loss of
generality, let us consider just one chromosome with $d$ base
pairs. Let $\mathbf x_1\in\RR^d, \dots, \mathbf x_n\in\RR^d$ be the
vectors of coverage across that chromosome. For example
$d=249,250,621$ is the number of base pairs on chr1, and $\mathbf
x_i\in\RR^d$ is the H3K4me3 coverage profile on chr1 for each sample
$i\in\{1, \dots, n\}$.

We also have exactly four sets of annotated regions $\underline R_i,
\overline R_i, R_i^+, R_i^-$ for each sample $i\in\{1, \dots, n\}$
(Table~\ref{tab:colors}). Each region $\mathbf r\in R_i^+$ is an
interval of base pairs. For example, Figure~\ref{fig:PeakError} shows
a coverage profile that has one annotation of each type.

\begin{table}[t!]
  \caption{Symbols and colors used to represent ChIP-seq 
    coverage and annotated regions for $n$ samples.}
  \label{tab:colors}
  \begin{center}
      \begin{tabular}{rccc}
  Data & Type & Color &Symbols \\
  \hline
  Coverage& &  & $\mathbf x_1, \dots \mathbf x_n$ \\
  Weak annotations & peaks
  & \annsquare{peaks}
  & $R_1^+, \dots, R_n^+$ \\
  & noPeaks 
  & \annsquare{noPeaks}
  & $R_1^-, \dots, R_n^-$ \\
  Strong annotations 
  & peakStart 
  & \annsquare{peakStart}
  & $\underline R_1, \dots, \underline R_n$ \\
  & peakEnd 
  & \annsquare{peakEnd}
  & $\overline R_1, \dots, \overline R_n$ \\
 \end{tabular}
  \end{center}
\end{table}

A peak detection function or peak caller $c:\RR^d \rightarrow \{0,
1\}^d$ takes a coverage profile $\mathbf x\in\RR^d $ as input, and
returns a binary peak call prediction $\mathbf y=c(\mathbf x)\in\{0,
1\}^d$ (0 is background noise, 1 is a peak).

The goal is to learn how to call peaks $c(\mathbf x_i)$ which agree
with the annotated regions $R_i^+, R_i^-, \underline R_i, \overline
R_i$ for some test samples $i$. To quantify the error of the peak
calls with respect to the annotation data, we define the annotation
error as the sum of false positive (FP) and false negative (FN)
regions:

\begin{equation}
  \label{eq:E}
  E(\mathbf y,  \underline R_i, \overline R_i, R_i^+, R_i^-) =
  \text{FP}(\mathbf y,  \underline R_i, \overline R_i, R_i^-) +
  \text{FN}(\mathbf y, \underline R_i, \overline R_i, R_i^+).
\end{equation}
The principle used to compute the annotation error $E$ is illustrated
in Figure~\ref{fig:PeakError}, and the precise mathematical
definitions of FP and FN are given in the supplementary
materials. In short, a false positive occurs when a peak detector
predicts too many peaks in an annotated region, and a false negative
occurs when there are not enough predicted peaks.

The supervised machine learning problem can be formalized as the
following optimization problem. Find the peak caller $c$ with minimal
annotation error on a set of test samples:
\begin{equation}
  \label{eq:min_ann_err}
  \minimize_c
  \sum_{i\in\text{test}}
  E\left[
    c(\mathbf x_i), \underline R_i, \overline R_i, R_i^+, R_i^-
  \right].
\end{equation}
In other words, the final goal is minimize the number of incorrect
annotated regions on a test set of data.
Note that ``test'' can be several different kinds of data sets,
depending on how the peak detection model will be used (see Results
section and Table~\ref{tab:train-test}).

\subsection{Calibrating a peak detection algorithm}

A standard unsupervised peak caller can be characterized as a function
$c_\lambda:\RR^d \rightarrow \{0, 1\}^d$, where the significance
threshold $\lambda\in\RR$ controls the number of peaks detected. In
each peak detection algorithm, $\lambda$ has a different, precise
meaning that we specify in the supplementary materials. As shown in
Figure~\ref{fig:train-error}, we select an optimal threshold $\lambda$
by minimizing the annotation error on the set of $n$ training samples
\begin{equation}
  \label{eq:min_err_peaks}
  \hat \lambda = \argmin_{\lambda}
  \sum_{i\in \{1, \dots, n\} }
  E\left[
    c_\lambda(\mathbf x_i), \underline R_i, \overline R_i, R_i^+, R_i^-
  \right].
\end{equation}
The training or model calibration procedure (\ref{eq:min_err_peaks})
consists of simply computing peak calls for several peak detection
thresholds $\lambda$, and choosing whichever threshold $\hat \lambda$
minimizes the number of incorrect annotated regions. Or if there are
no annotated regions available, we can simply use the default
significance threshold $\tilde \lambda$ suggested by the author of
each algorithm. The test error (\ref{eq:min_ann_err}) can be used to
evaluate the accuracy of the trained model $\hat \lambda$ and the
default model $\tilde \lambda$.

\subsection{Specific peak detection algorithms and parameters}

We considered the following algorithms with free software
implementations from the bioinformatics literature. We chose to
compare several algorithms designed for peaks (such as H3K4me3) and
several algorithms for broad domains (such as H3K36me3). However, our
analysis could be conducted with any peak detection algorithm, and we
hope that authors of future algorithms will test them using annotated
region databases.

\begin{figure}[b!]
\begin{center}
  \includegraphics[width=0.5\linewidth]{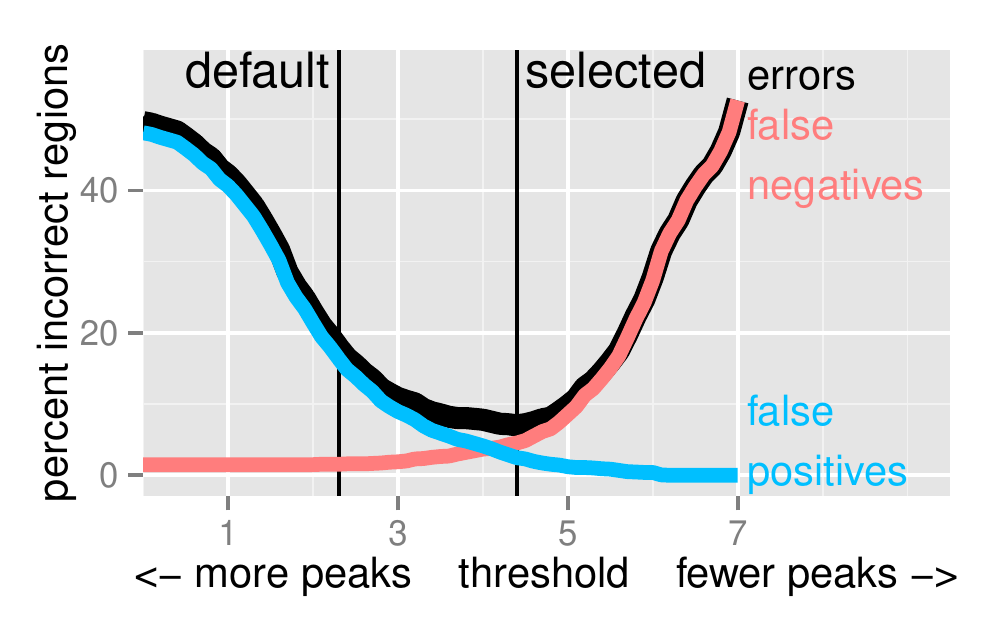}
\end{center}
\vskip -0.7cm
\caption{The annotated regions define an annotation error function
  that can be minimized (\ref{eq:min_err_peaks}) to choose an optimal
  parameter for a standard peak detector. We plot the percent error of
  the hmcan.broad model over 1743 annotated regions in the
  H3K36me3\_AM\_immune data set (other algorithms and data sets shown
  in \figtrain).}
\label{fig:train-error}
\end{figure}

MACS and HMCan \citep{MACS, HMCan} have parameters for peaks and broad
domains, so we tried both settings.  RSEG and SICER were designed for
broad domains \citep{RSEG, SICER}. The HOMER set of tools contains a
findPeaks program which has been used to detect transription factor
binding sites and histone modifications \citep{homer}. Details of each
algorithm are given in the supplementary materials.

Each algorithm has several parameters that may affect peak detection
accuracy. An exhaustive grid search over several parameters would be
infeasible, since there is an exponential number of different
parameter combinations. So for each algorithm we use only one
parameter as the significance threshold $\lambda$, and held other
parameters at default values. The precise meaning of the threshold
$\lambda$ for each algorithm is discussed in the supplementary
materials.

\begin{figure*}[p]
\begin{center}
  \includegraphics[width=\linewidth]{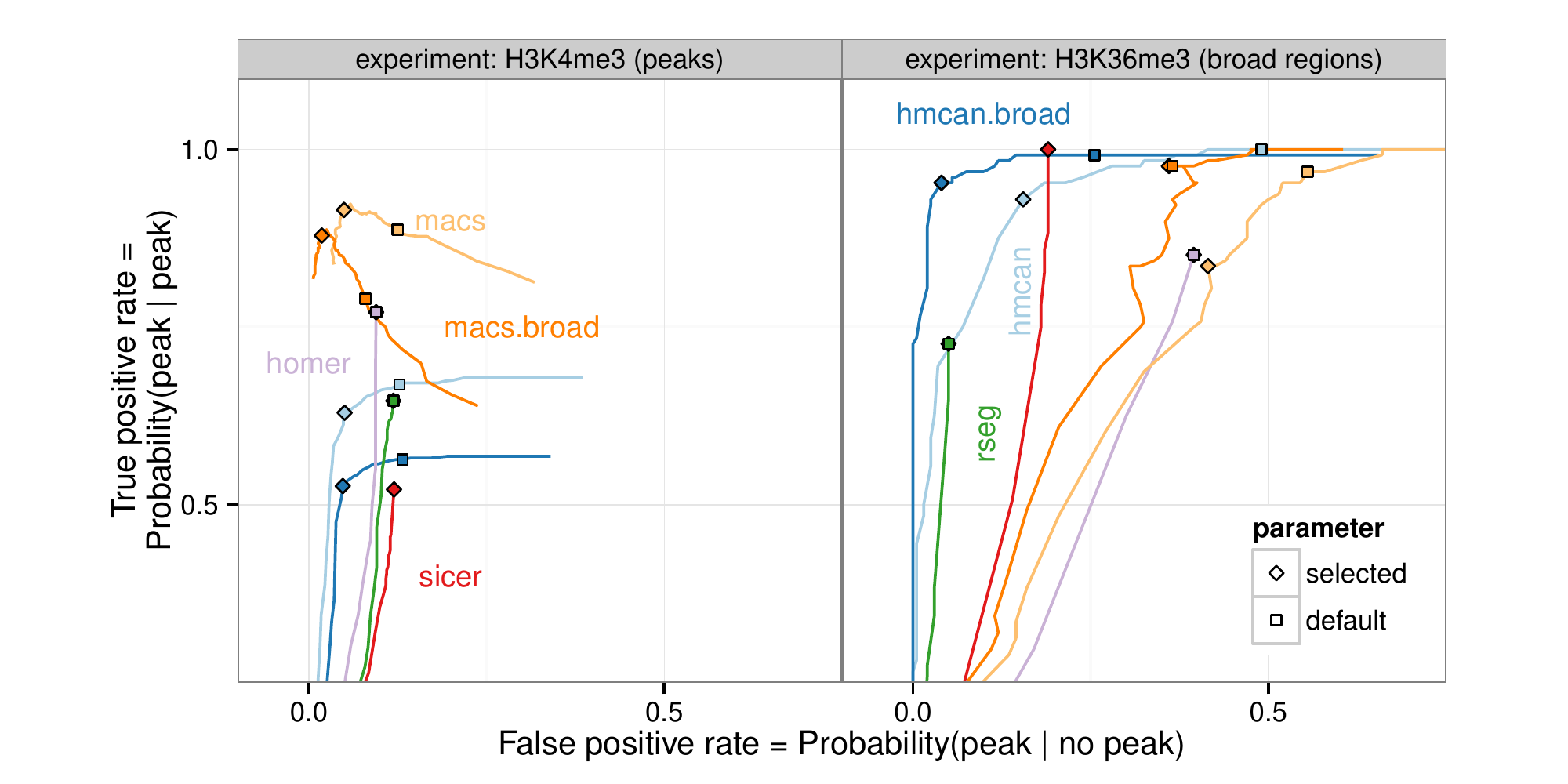}
\end{center}
\vskip -0.7cm
\caption{ROC curves show that different peak detectors are better for
  different mark types. We show data for annotator TDH, and the same
  model ordering was observed across the other 3 annotators
  (\figseveral). Note that even though we included all possible peaks
  (no threshold parameter $\lambda=0$), some algorithms such as rseg
  and homer do not come close to 100\% true positive rates. Also, some
  algorithms such as macs and macs.broad show non-monotonic ROC curves
  since their qvalue threshold parameter $\lambda$ changes the size of
  each peak, in addition to the number of peaks.}
\label{fig:roc}
\end{figure*}

\begin{figure*}[p]
\begin{center}
  \includegraphics[width=\linewidth]{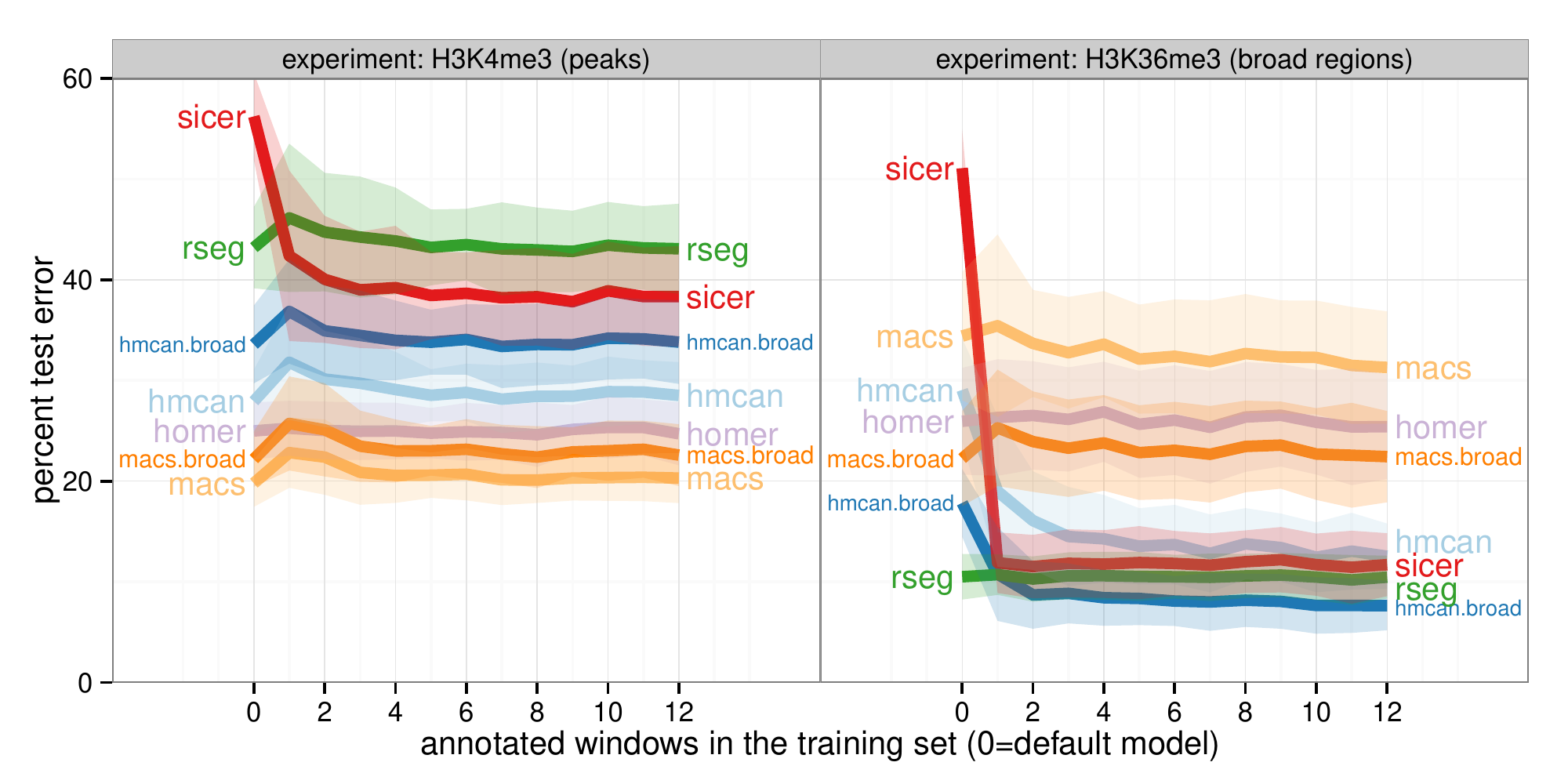}
\end{center}
\vskip -0.7cm
\caption{Test error curves show mean and standard deviation over 100
  randomly selected train and test sets, using the data sets with the
  most annotated windows for each mark type: H3K4me3\_PGP\_immune and
  H3K36me3\_AM\_immune (Table~\ref{tab:counts}). Differences between
  models are clearly visible even in train data sets with just a few
  annotated windows. Similar model orderings were observed in other
  train/test splits (\figannotators\ and \figtypes).}
\label{fig:test-same}
\end{figure*}

\subsection{Software availability}

The annotation error (\ref{eq:E}) can be easily computed using the C
code in the R
package PeakError on GitHub:\\
\url{https://github.com/tdhock/PeakError}\\
The \verb|PeakError| function takes as input the annotated regions and
peak predictions for one sample, and outputs the annotation error for
each region.

\section{Results}


\subsection{Creating a benchmark data set of annotated regions}

We analyzed samples from the McGill Epigenomes portal
(\url{http://epigenomesportal.ca}), which we visualized using the UCSC
genome browser software \citep{ucsc}. We chose to annotate all samples
available with data on H3K4me3 (representing profiles with narrow
peaks) and all samples available with data on H3K36me3 (representing
broadly enriched regions). In these data there are many samples of the
same cell type, and the same peaks often occur in all samples of the
same cell type (Figure~\ref{fig:annotations}). In total there were 37
samples with H3K4me3 data and 29 samples with H3K36me3 data, across 8
cell types (Table~\ref{tab:counts}).

We constructed 7 annotated region databases (Table~\ref{tab:counts}),
using the method described in the ``Annotating samples'' section. Of
the 4 different annotators, some were post-docs (TDH, XJ), and some
were PhD students (AM, PGP). In total we created 12,826 annotated
regions.

The annotated region databases should be useful for benchmarking
future peak detection algorithms. They can be viewed and downloaded at
\url{http://cbio.ensmp.fr/~thocking/chip-seq-chunk-db/}

\subsection{Train error for different mark types and annotators}

We considered annotation data sets of two different histone mark
types: H3K4me3 and H3K36me3. Is there any single algorithm that can
fit data from both mark types? Or is each algorithm better for fitting
one mark type or the other?

We first considered training each peak detection algorithm by choosing
the threshold with minimum annotation error $\hat \lambda$
(\ref{eq:min_err_peaks}) on each data set (\figtrain). The macs
algorithm showed the minimum train error of 8.9--20\% across the 4
annotated H3K4me3 data sets. The hmcan.broad algorithm showed the
minimum train error of 7--20.2\% across the 3 annotated H3K36me3 data
sets.

We also computed ROC curves by varying the peak detection threshold
$\lambda$ for every algorithm (Figure~\ref{fig:roc}). The ROC curves
show that different algorithms are good detectors for
different mark types. For example, it is clear that macs was the best
peak detector for H3K4me3, and hmcan.broad was the best peak detector
for H3K36me3. We observed similar trends in the ROC curves for the
other annotation data sets (\figroc).

\begin{table}[t!]
  \caption{Examples of different train and test sets. 
    Note that the histone mark type is always the same 
    (for example, we did not try to train a model on H3K4me3 
    and test it on H3K36me3).}
  \label{tab:train-test}
  \begin{center}
      \begin{tabular}{ccccc}
    Same & 
    Train & Test \\
    \hline
    annotator, cell types & 
    chr1 & chr2 \\
    annotator, chromosomes & 
    T cells & kidney cells \\
    cell types, chromosomes & 
    annotator TDH & annotator PGP
  \end{tabular}
  \end{center}
\end{table}

The ROC curves and the annotation error curves also show that the
default parameter values $\tilde \lambda$ have generally higher false
positive rates than the optimal parameter values $\hat \lambda$ chosen
by minimizing the annotation error (\ref{eq:min_err_peaks}).  For
example, \figlearned\ shows two profiles where the default algorithms
show false positives, but the trained algorithms achieve perfect peak
detection.

\begin{table*}[t!]
  \caption{Counts of windows, annotated regions,
    and samples of each cell type
    in each of the 7 data sets we analyzed.}
  \label{tab:counts}
  \begin{center}
    \begin{scriptsize}
\begin{tabular}{rrrrrrrr}
  
  \hline
mark & H3K36me3 & H3K4me3 & H3K36me3 & H3K36me3 & H3K4me3 & H3K4me3 & H3K4me3 \\ 
  annotator & TDH & TDH & AM & TDH & PGP & TDH & XJ \\ 
  sample.set & other & other & immune & immune & immune & immune & immune \\ 
  \hline
windows & 4 & 29 & 23 & 4 & 30 & 27 & 12 \\ 
  \hline
noPeaks & 72 & 536 & 752 & 230 & 1653 & 1656 & 702 \\ 
  peakStart & 68 & 305 & 294 & 200 &  813 &  796 & 216 \\ 
  peakEnd & 60 & 311 & 294 & 200 &  730 &  933 & 216 \\ 
  peaks &  & 218 & 403 &  &  638 &  287 & 243 \\ 
  \hline
tcell &  &  &  15 &  15 &   19 &   19 &  19 \\ 
  monocyte &  &  &   5 &   5 &    6 &    6 &   6 \\ 
  bcell &  &  &   1 &   1 &    2 &    2 &   2 \\ 
  kidney &  1 &   1 &  &  &  &  &  \\ 
  kidneyCancer &  1 &   1 &  &  &  &  &  \\ 
  skeletalMuscleCtrl &  3 &   3 &  &  &  &  &  \\ 
  skeletalMuscleMD &  3 &   4 &  &  &  &  &  \\ 
  leukemiaCD19CD10BCells &  &   1 &  &  &  &  &  \\ 
   \hline
\end{tabular}

    \end{scriptsize}
  \end{center}
\end{table*}

\subsection{Consistency of different annotators}

When two experts annotate the same profiles, are they consistent?
Does a peak detector trained on one annotator work well for another
annotator? To answer these questions, we asked 4 different people to
create annotated regions based on the same ChIP-seq profiles.

There were some windows in the H3K36me3 immune data sets that were
annotated by both TDH and AM (\figseveral). It is clear that these two
annotators had consistent definitions of peak locations, but different
levels of detail. For example, TDH used a peakStart and peakEnd region
in many instances where AM used a peaks region. Also, there are other
regions which only TDH annotated, and AM left un-annotated. 

When we looked at train error of the different algorithms, we saw the
same mark-specific model ordering that was independent of the
annotator. For example, \figroc\ shows similar ROC curves for the
H3K4me3 immune annotation data sets, across 3 different annotators
(TDH, PGP, and XJ). In particular, it is clear that macs is the best
algorithm across the 3 annotators.

We also observed similar model orderings across annotators in
terms of test error (\figannotators). In particular, we trained models
on annotations from TDH, and tested them on annotations from PGP
(holding the mark H3K4me3 and immune cell types constant). We observed
very little changes in test error when training on one or the other
annotator.

Overall, these data indicate that annotated regions were consistent
across annotators. Annotated regions can thus be used as a method for
benchmarking peak detection algorithms, with results that are specific
to the ChIP-seq data set, and quite robust across annotators.

\subsection{Test on different regions of the same samples}

Does a peak detector trained on just a few annotated chromosomes work
well for detecting peaks on the other chromosomes of the same samples?

To answer this question, we performed cross-validation on each
annotation data set by designating half of the windows as a test
set. Then, we trained a model using only a few of the other windows,
for training set sizes of from 1 to 12 windows. We randomly selected
100 train and test sets, trained each peak detector by picking the
parameter $\hat \lambda$ with minimal error on the training set
(\ref{eq:min_err_peaks}), then evaluated its peak detection by
computing the annotation error on the test set of windows
(\ref{eq:min_ann_err}).

Figure~\ref{fig:test-same} shows the test error as a function of
training data size for two annotation data sets. We observed that each
algorithm quickly achieves its model-specific minimum error, after
only about 4 annotated windows in the train set. Furthermore, we
observed that macs had the lowest test error for H3K4me3 data sets,
and hmcan.broad had the lowest test error for H3K36me3 data sets.

Figure~\ref{fig:test-same} also shows the test error of the default
thresholds $\tilde \lambda$ suggested by the author of each
algorithm. For some models such as sicer, the trained model $\hat
\lambda$ is clearly better than the default model $\tilde
\lambda$. For other models such as rseg, the trained model $\hat
\lambda$ has about the same test error as the default model $\tilde
\lambda$.

\subsection{Test on different samples and cell types}

Does a peak detector trained on some cell types work when applied to
other cell types? 

To answer this question, we trained each algorithm on several immune
cell types (tcell, bcell, monocyte) and tested them on other cell
types (kidney, kidney cancer, skeletal muscle, leukemia). We observed
that most algorithms had similar test error when training on the same
or different cell types (\figtypes). The exceptions were the macs and
macs.broad algorithms, which exhibited higher test error when training
on samples of different cell types. These data suggest different
samples have different kinds of background noise, and that it is
important to annotate all samples of interest for accurate peak
detection using the macs algorithm.

\section{Discussion}

\subsection{Supervised versus unsupervised analysis}

In the machine learning literature, a problem is considered supervised
when there is a teacher or expert that provides correct predictions
for a learning algorithm. In this paper, the type of supervision that
we proposed was a database of annotated regions that represents where
a scientist does and does not observe peaks. We used these annotated
regions as a gold standard to define a supervised learning problem
(\ref{eq:min_ann_err}), which seeks a peak detector with minimal
incorrect regions on a test data set. Furthermore, we proposed to
train or calibrate standard ChIP-seq peak callers by choosing a
threshold $\hat \lambda$ (\ref{eq:min_err_peaks}) that minimizes the
number of incorrect regions.

In contrast, ChIP-seq peak detection without annotated regions can be
considered an unsupervised learning problem. Usually, a default
threshold $\tilde \lambda$ and peak detection algorithm are first fit
to a data set, and then the peaks are judged by visualizing them along
with the data in a genome browser. If there are too many peaks the
user can increase the threshold $\lambda$, or if there are not enough
peaks the user can decrease the threshold $\lambda$.

In fact, our supervised analysis protocol with annotated regions is
very similar, but is independent of any specific peak detection
algorithm. First, one must visually inspect the ChIP-seq data in a
genome browser, and create an annotated region database. After that,
those annotations can be used to choose an appropriate model, and to
calibrate its parameters.

\subsection{Time required for annotation}

Annotated regions are applicable to any ChIP-seq data analysis
project, but their main weakness is the time it takes to create the
annotation database.  However, we found that it only takes about 10
minutes of manual visual inspection to find and annotate a whole
window of several nearby regions. 

Additionally, we were able to quickly create many regions by
annotating dozens of samples at the same time. For example, the immune
H3K4me3 sample set consists of 27 
samples (Table~\ref{tab:counts}). So when we found a
region with a peak across all samples, we assigned the same peaks
annotated region to all 27 samples.

As another example, it only took about 40 minutes to create the
H3K36me3\_TDH\_other data set, which contains in total 8 samples and
200 annotated regions across 4 genomic windows
(Table~\ref{tab:counts}).
And even though this was the smallest data
set that we created, we were still able to observe clear differences
in train error between the various algorithms (\figtrain\ and
\figroc).

Also, the test error curves indicate that only a few annotated windows
are necessary to calibrate peak detection thresholds $\lambda$
(Figure~\ref{fig:test-same}). In both H3K4me3 and H3K36me3 data sets,
the test error decreases to its model-specific minimum after
annotating only about 4 windows.

Overall, these data indicate that a relatively small annotated region
database of a few annotated windows across several samples is
sufficient to calibrate and test peak detection algorithms.

\section{Conclusions}

We propose a supervised machine learning approach to ChIP-seq data
analysis. Our approach involves first creating an annotated region
database for a specific data set, and then using it to choose an
appropriate peak detection algorithm.

We used this approach to benchmark the performance of several peak
detectors on several H3K4me3 and H3K36me3 data sets. We observed that
macs was the best peak detector for H3K4me3 data, and hmcan.broad was
the best for H3K36me3 data. Furthermore, we observed these trends in
train and test error, across several different annotators and cell
types. 

However, even the best peak detectors exhibited 10--20\% test error
(Figure~\ref{fig:test-same}), which could be improved. To develop even
better peak detectors, we are interested in developing multi-parameter
supervised learning algorithms, which have proven to be very
successful in breakpoint detection \citep{HOCKING-penalties}. 

We have made our annotated regions available as a public benchmark
data set. Such annotated data sets are essential for making links
between computational biology and the larger computer science research
community. In particular, these annotated data sets make the ChIP-seq
peak detection problem more accessible to machine learning
researchers, who will now be able to work on developing supervised
learning algorithms for peak detection.
For example, new algorithms for broad marks should be tested
alongside the hmcan.broad model, which we observed to be the best for
H3K36me3 data. In the future, annotated region databases should be
created to benchmark algorithms for other ChIP-seq experiments
(e.g. H3K9me3).

We are interested in using systems such as Apollo \citep{apollo},
which extend genome browsers to support interactive annotation. We
would also like to develop an interactive ChIP-seq data analysis
system that displays optimal peak predictions which can be updated by
adding annotated regions.

\section{Acknowledgements}

Thanks to David Bujold for help with the Epigenomes portal data and
thanks to Haitham Ashoor, Chris Benner, Maxime Caron, Tao Liu, and
Song Qiang for help with the peak callers.

\subsection{Funding} 
This work was supported by computing resources provided by Calcul
Quebec and Compute Canada, and by Canadian Institutes of Health
Research grants EP1-120608 and EP1-120609, awarded to GB.
\subsection{Conflict of interest statement.} None declared.

\newpage

\begin{center}
  All of the sections after this page are to be considered
  supplementary materials.
\end{center}

\appendix

\newpage

\section{Source code and data links}

\subsection{Original ChIP-seq data profiles}

\url{http://epigenomesportal.ca}

\subsection{Annotated region database}

\url{http://cbio.ensmp.fr/~thocking/chip-seq-chunk-db/}

\subsection{R package with C code for computing the annotation error}

\url{https://github.com/tdhock/PeakError}

\subsection{\LaTeX\ and R source code for this article}

\url{https://bitbucket.org/mugqic/chip-seq-paper}

\section{Example of annotated region data file}

Below is an example of an annotation file that TDH created by manual
visual inspection of the H3K4me3 immune cell samples (tcell, bcell, and
monocyte). It is divided into 4 windows, each of which is separated by
two returns/newlines. Each line represents an annotated region for
several cell types. The cell types that are not listed all get a
noPeaks annotation in the indicated region. Genomic regions were
copied from the UCSC genome browser web page, and pasted into a text
file.

\begin{verbatim}
chr11:118,092,641-118,095,026 peakStart monocyte tcell
chr11:118,095,334-118,096,640 peakEnd monocyte tcell
chr11:118,101,452-118,118,472 peaks
chr11:118,121,649-118,124,175 peaks monocyte tcell bcell

chr11:111,285,081-111,285,355 peakStart monocyte
chr11:111,285,387-111,285,628 peakEnd monocyte
chr11:111,299,681-111,337,593 peaks
chr11:111,635,157-111,636,484 peakStart monocyte tcell bcell
chr11:111,637,473-111,638,581 peakEnd monocyte tcell bcell

chr1:32,717,194-32,721,976 peaks tcell
chr1:32,750,608-32,756,699 peaks
chr1:32,757,261-32,758,801 peaks tcell bcell monocyte

chr2:26,567,544-26,568,406 peakStart bcell tcell monocyte
chr2:26,568,616-26,568,862 peakEnd bcell tcell monocyte
chr2:26,569,573-26,571,905 peakEnd bcell tcell monocyte
chr2:26,578,595-26,632,223 peaks
chr2:26,634,282-26,636,118 peaks monocyte
\end{verbatim}

\newpage \section{Definition of the annotation error}

In this section we give the precise mathematical definition of the
annotation error.

\subsection{Data definition}

Let there be $n$ annotated training samples, all of the same histone
mark type. For simplicity, and without loss of generality, let us
consider just one chromosome with $d$ base pairs. Let $\mathbf
x_1\in\RR^d, \dots, \mathbf x_n\in\RR^d$ be the vectors of coverage
across that chromosome. For example $d=249,250,621$ is the number of
base pairs on chr1, and $\mathbf x_i\in\RR^d$ is the H3K4me3 coverage
profile on chr1 for one sample $i\in\{1, \dots, n\}$.

We also have exactly 4 sets of annotated regions $\underline R_i,
\overline R_i, R_i^+, R_i^-)$ for each sample $i\in\{1, \dots, n\}$:
\begin{center}
  \begin{tabular}{rccc}
  Data & Type & Color &Symbols \\
  \hline
  Coverage& &  & $\mathbf x_1, \dots \mathbf x_n$ \\
  Weak annotations & peaks
  & \annsquare{peaks}
  & $R_1^+, \dots, R_n^+$ \\
  & noPeaks 
  & \annsquare{noPeaks}
  & $R_1^-, \dots, R_n^-$ \\
  Strong annotations 
  & peakStart 
  & \annsquare{peakStart}
  & $\underline R_1, \dots, \underline R_n$ \\
  & peakEnd 
  & \annsquare{peakEnd}
  & $\overline R_1, \dots, \overline R_n$ \\
 \end{tabular}
\end{center}
For each sample $i\in\{1, \dots, n\}$, $R_i^+$ is a set of regions,
and each region is an interval of base pairs. For example,
Figure~\ref{fig:PeakError} shows a sample with one annotation of each
type.

A peak detection function or peak caller $c:\RR^d \rightarrow \{0, 1\}^d$ 
gives a binary peak call prediction $\mathbf y=c(\mathbf x)\in\{0,
1\}^d$ given some coverage profile $\mathbf x$.

The goal is to learn how to call peaks $c(\mathbf x_i)$ which agree
with the annotated regions $R_i^+, R_i^-, \underline R_i, \overline
R_i$ for some test samples $i$. To quantify the error of the peak
calls with respect to the annotation data, we define the following
functions.

\subsection{Weak annotation error}

The weak annotations $R_i^+,R_i^-$ count whether there are any peaks
overlapping a region. They are called weak because each peaks region
can only produce a false negative (but not a false positive), and each
noPeaks region can only produce a false positive (but not a false
negative). Let
\begin{equation}
  B(\mathbf y, \mathbf r) = \sum_{j\in \mathbf r} y_j
\end{equation}
be the number of bases which have peaks overlapping region $\mathbf
r$. Then for a sample $i$, the number of weak false positives is
\begin{equation}
  \label{eq:WFP}
  \text{WFP}(\mathbf y, R_i^-) = 
  \sum_{\mathbf r \in R_i^-} 
  I\left[ B(\mathbf y, \mathbf r) > 0 \right]
\end{equation}
and the number of weak true positives is
\begin{equation}
  \label{eq:WTP}
  \text{WTP}(\mathbf y, R_i^+) = 
  \sum_{\mathbf r \in R_i^+} 
  I\left[ B(\mathbf y, \mathbf r) > 0 \right],
\end{equation}
where $I$ is the indicator function.

\subsection{Strong annotation error}

The strong annotations $\underline R_i, \overline R_i$ count the
number of peak starts and ends occuring in the given regions. They are
called strong because each region can produce a false positive (more
than 1 peak start/end predicted in the region) or a false negative (no
peak start/end predicted). First, let $y_0 = y_{d+1} = 0$ and define
the set of first bases of all peaks as
\begin{equation}
  \label{eq:underI}
  \underline {\mathcal I}(\mathbf y) = \left\{
  j\in\{1, \dots, d\}: y_j=1 \text{ and } y_{j-1}=0
  \right\}
\end{equation}
and the set of last bases of all peaks as
\begin{equation}
  \label{eq:overI}
  \overline {\mathcal I}(\mathbf y) = \left\{
  j\in\{1, \dots, d\}: y_{j}=1\text{ and } y_{j+1}=0
  \right\}.
\end{equation}

For a sample $i$, the number of strong false positives is
\begin{equation}
  \label{eq:SFP}
  \text{SFP}(\mathbf y, \underline R_i, \overline R_i) =
  \sum_{\mathbf r\in\underline R_i}
  I\big[ |\mathbf r\cap \underline {\mathcal I}(\mathbf y)| > 1 \big] +
  \sum_{\mathbf r\in\overline R_i}
  I\big[ |\mathbf r\cap \overline {\mathcal I}(\mathbf y)| > 1 \big], 
\end{equation}
and the number of strong true positives is
\begin{equation}
  \label{eq:STP}
  \text{STP}(\mathbf y, \underline R_i, \overline R_i) =
  \sum_{\mathbf r\in\underline R_i}
  I\big[ |\mathbf r\cap \underline {\mathcal I}(\mathbf y)| > 0 \big] +
  \sum_{\mathbf r\in\overline R_i}
  I\big[ |\mathbf r\cap \overline {\mathcal I}(\mathbf y)| > 0 \big].
\end{equation}

\subsection{Total annotation error}

For a sample $i$, the total number of false positives is
\begin{equation}
  \label{eq:FP}
  \text{FP}(\mathbf y,  \underline R_i, \overline R_i, R_i^-) =
  \text{WFP}(\mathbf y, R_i^-)+
  \text{SFP}(\mathbf y, \underline R_i, \overline R_i),
\end{equation}
the total number of true positives is
\begin{equation}
  \label{eq:FN}
  \text{TP}(\mathbf y, \underline R_i, \overline R_i, R_i^+) =
  \text{WTP}(\mathbf y, R_i^+) +
  \text{STP}(\mathbf y, \underline R_i, \overline R_i),
\end{equation}
the total number of false negatives is
\begin{equation}
  \label{eq:FN}
  \text{FN}(\mathbf y, \underline R_i, \overline R_i, R_i^+) =
  |\underline R_i| + |\overline R_i| + |R_i^+|
  -\text{TP}(\mathbf y, \underline R_i, \overline R_i, R_i^+).
\end{equation}
The annotation error $E$
quantifies the number of incorrect annotated regions:
\begin{equation}
  \label{eq:E}
  E(\mathbf y,  \underline R_i, \overline R_i, R_i^+, R_i^-) =
  \text{FP}(\mathbf y,  \underline R_i, \overline R_i, R_i^-) +
  \text{FN}(\mathbf y, \underline R_i, \overline R_i, R_i^+).
\end{equation}
The annotation error $E$ can be easily computed using the C code
in the R package PeakError on GitHub:
\url{https://github.com/tdhock/PeakError}.

\subsection{ROC analysis}

A standard unsupervised peak caller can be characterized as a function
$c_\lambda:\RR^d \rightarrow \{0, 1\}^d$, where the significance
threshold $\lambda\in\RR$ controls the number of peaks detected. 
Receiver Operating Characteristic (ROC) curves can also be used to
show the train error of all thresholds of a peak detector on an
annotation data set. Define the false positive rate as
\begin{equation}
  \label{eq:FPR}
  \text{FPR}(\lambda) =
  \frac{
    \sum_{i=1}^n
    \text{FP}\left[
      c_\lambda(\mathbf x_i),  \underline R_i, \overline R_i, R_i^-
    \right]
  }{
   |\underline R_i| + |\overline R_i| + |R_i^-| 
 }
\end{equation}
and the true positive rate as
\begin{equation}
  \label{eq:TPR}
  \text{TPR}(\lambda) =
  \frac{
    \sum_{i=1}^n
    \text{TP}\left[ c_\lambda(\mathbf x_i), \underline R_i, \overline R_i, R_i^+
    \right]
  }{
    |\underline R_i| + |\overline R_i| + |R_i^+|
  }.
\end{equation}

ROC curves are traced by plotting $\text{TPR}(\lambda)$ versus
$\text{FPR}(\lambda)$ for all possible values of the peak detection
threshold $\lambda$.

\newpage \section{Details of peak calling algorithms}

\citet{MACS} proposed Model-based Analysis of ChIP-Seq (MACS). We
downloaded MACS version 2.0.10 12162013 (commit
ca806538118a85ec338674627f0ac53ea17877d9 on GitHub). The significance
threshold $\lambda$ is the qvalue cutoff parameter \texttt{-q}. We
used a grid of 59 qvalue cutoffs from 0.8 to $10^{-15}$. 

The macs broad peak caller is more difficult to train because it has 2
different qvalue cutoff parameters (\texttt{-q} and
\texttt{--broad-cutoff}). So we used the same grid of 59 qvalue
cutoffs for \texttt{-q}, and then defined
\begin{equation}
  \label{eq:broad-cutoff}
  \texttt{broad-cutoff} = \texttt{q} \times 1.122.
\end{equation}


The default macs algorithms use a default
qvalue cutoff of $\tilde \lambda = 0.05$.


\citet{HMCan} proposed the HMCan algorithm which uses GC-content and
copy number normalization. We downloaded HMCan commit
9d0a330d0a873a32b9c4fa72c94d00968132b9ef from BitBucket. We used the
default GC content normalization file provided by the authors. We used
two different parameter files to test two different peak detectors: 
\begin{center}
  \begin{tabular}{cr}
  name & mergeDistance \\
  \hline
  hmcan & 200 \\
  hmcan.broad & 1000
\end{tabular}
\end{center}
We then ran HMCan with finalThreshold=0, and
defined $\lambda$ as a threshold on column 5 in the
regions.bed file. Both hmcan and hmcan.broad use a default finalThreshold of
$\tilde \lambda = 10$.

\citet{RSEG} proposed the RSEG algorithm. We downloaded RSEG version
0.4.8 from \url{http://smithlabresearch.org/software/rseg/}. Upon
recommendation of the authors, we saved computation time by running
\texttt{rseg-diff} using options \texttt{-training-size 100000} and
\texttt{-i 20}. We used the \texttt{-d} option to specify a dead
regions file for hg19 based on our alignment pipeline. We defined the
significance threshold $\lambda$ as the sum of posterior scores
(column 6 in the output .bed file). For the default RSEG algorithm, we
used all the peaks in the output file, meaning a posterior score
threshold of $\tilde \lambda = 0$.

\citet{SICER} proposed the SICER algorithm. We downloaded SICER
version 1.1 from \url{http://home.gwu.edu/~wpeng/SICER\_V1.1.tgz}. We
defined the significance threshold $\lambda$ as the FDR (column 8) in
the islands-summary output file. For the default SICER algorithm, we
used an FDR of $\tilde \lambda=0.01$ as suggested in the README and
example.

\citet{homer} proposed the HOMER set of tools for DNA motif
detection. We used the \texttt{findPeaks} program in HOMER version 4.1
with the \texttt{-style histone} option. We defined the significance
threshold $\lambda$ as the ``p-value vs Control'' column. We defined
the default model as all peaks in the output file.

\newpage

\section{Supplementary Figures}

 \section*{\figtrain: annotation error curves for all
  algorithms and data sets}

\begin{center}
  \includegraphics[width=0.95\textwidth]{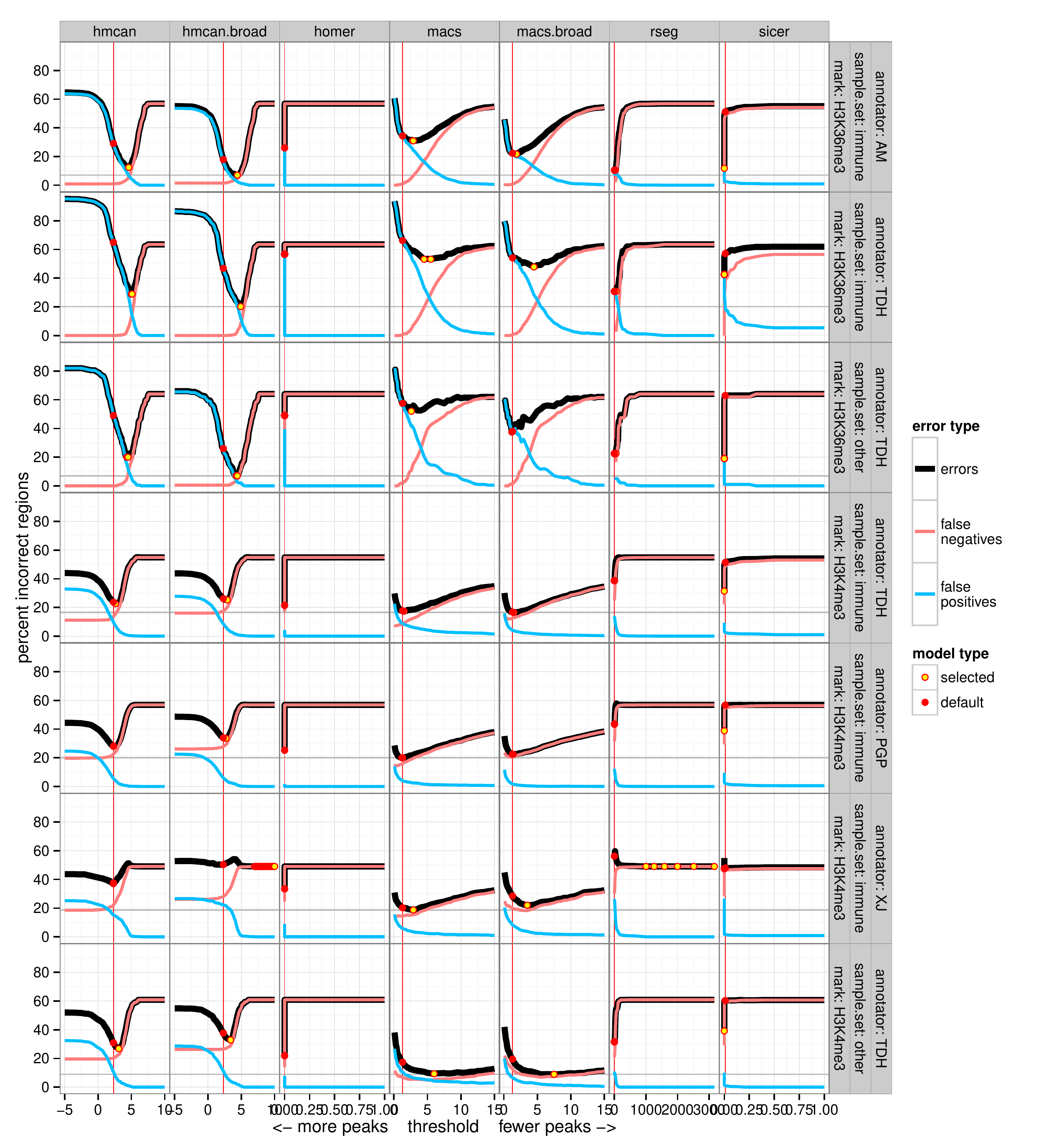}
\end{center}
Train error curves for each data set (rows) and each algorithm
(columns). A vertical red line shows the default parameter for each
model, and a horizontal grey line shows the minimum error model for
each data set.

\newpage \section*{\figroc: ROC curves for train error of all
  algorithms on all data sets}

\begin{center}
  \includegraphics[width=\textwidth]{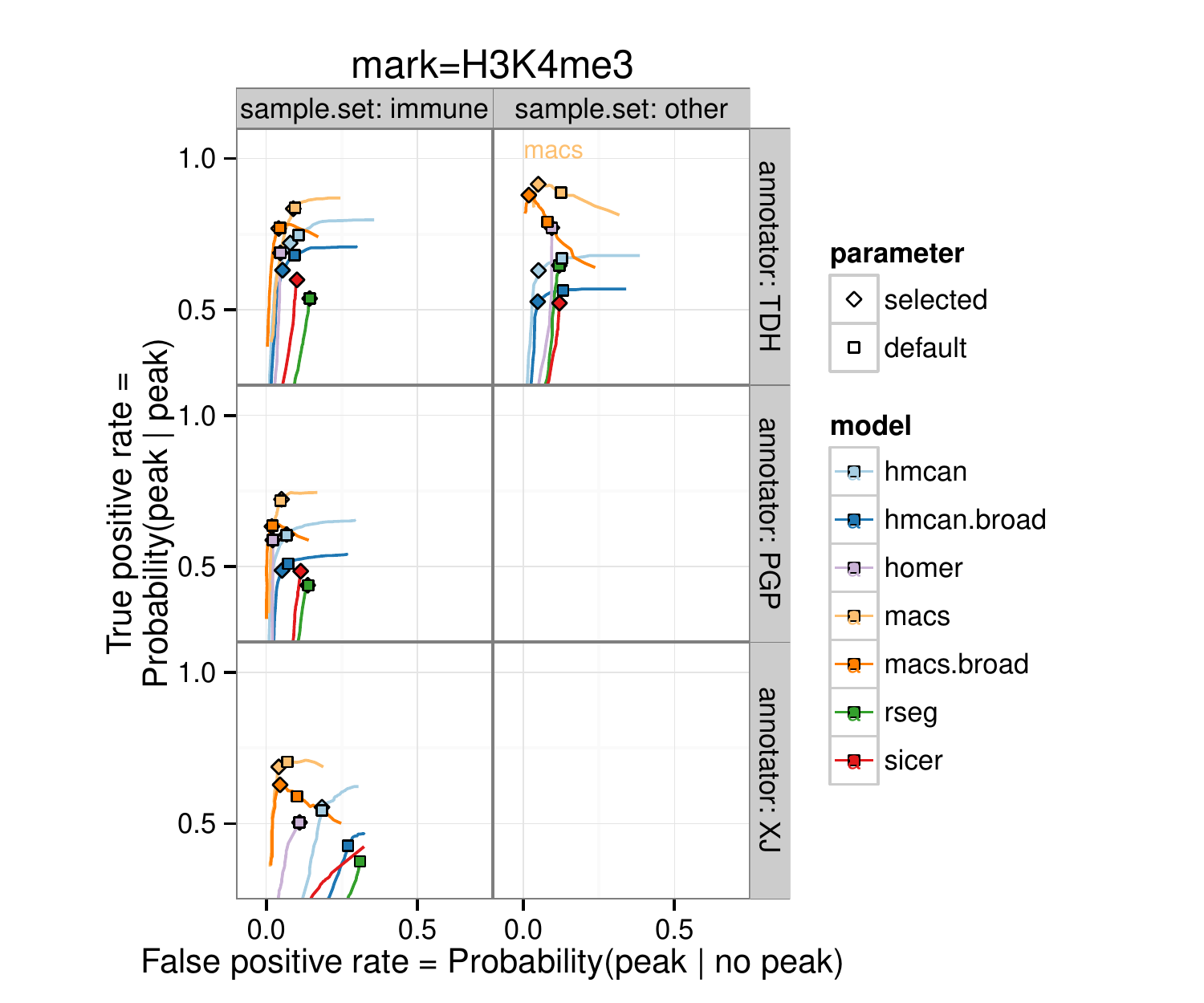}
\end{center}

It is clear that macs is the best for H3K4me3 data.

\begin{center}
  \includegraphics[width=\textwidth]{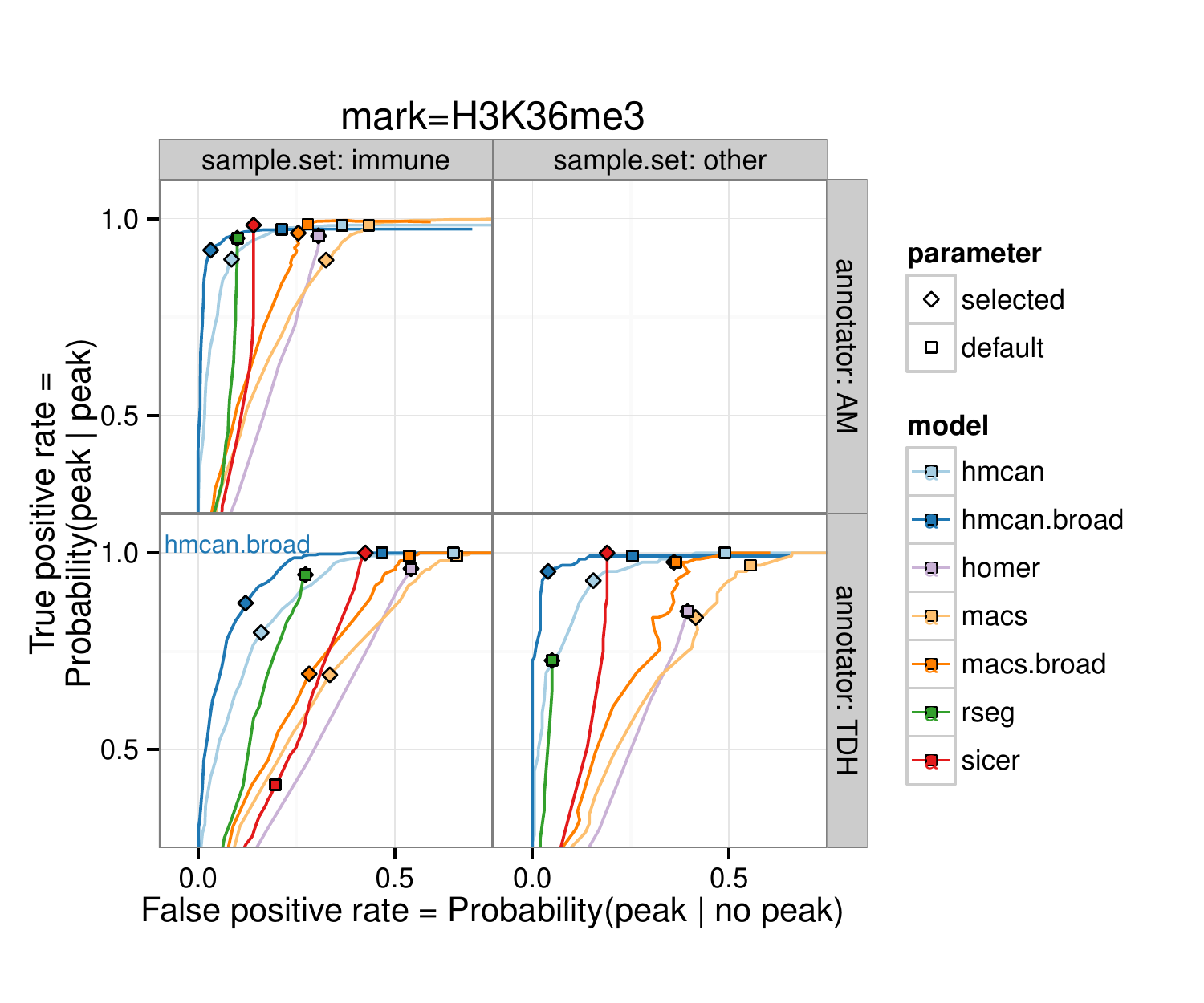}
\end{center}

It is clear that hmcan.broad is the best for H3K36me3 data.

\newpage \section*{\figlearned: examples of fitted peak detectors}

\begin{center}
  \includegraphics[width=\textwidth]{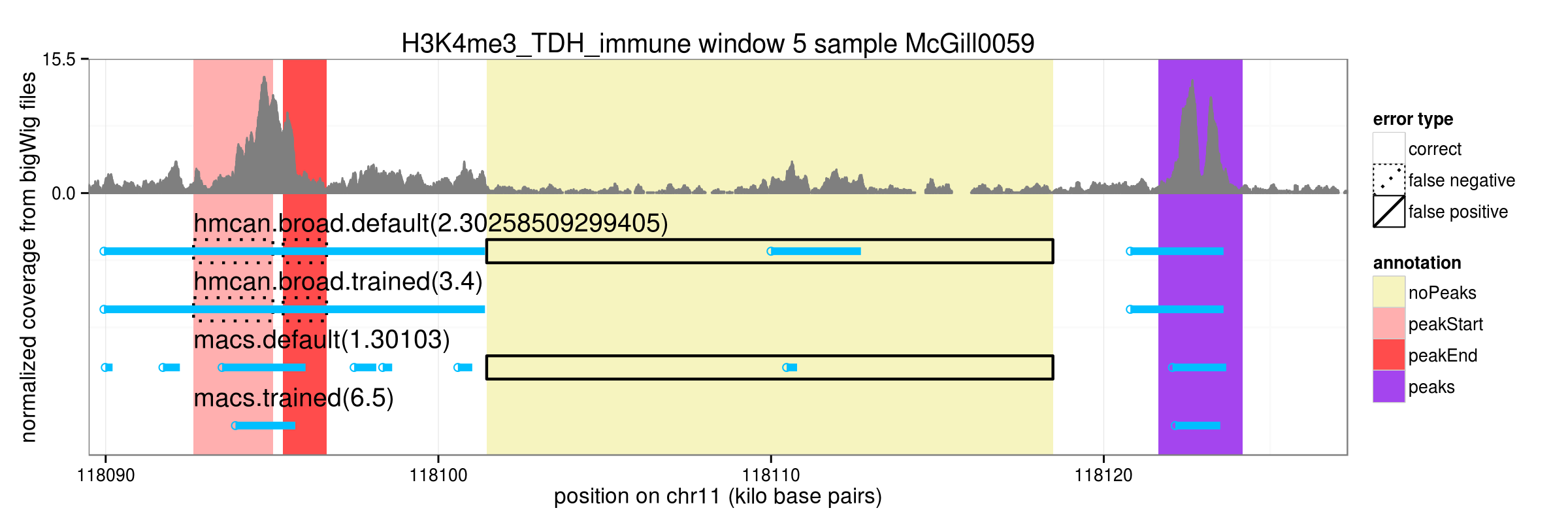}
\end{center}

\begin{center}
  \includegraphics[width=\textwidth]{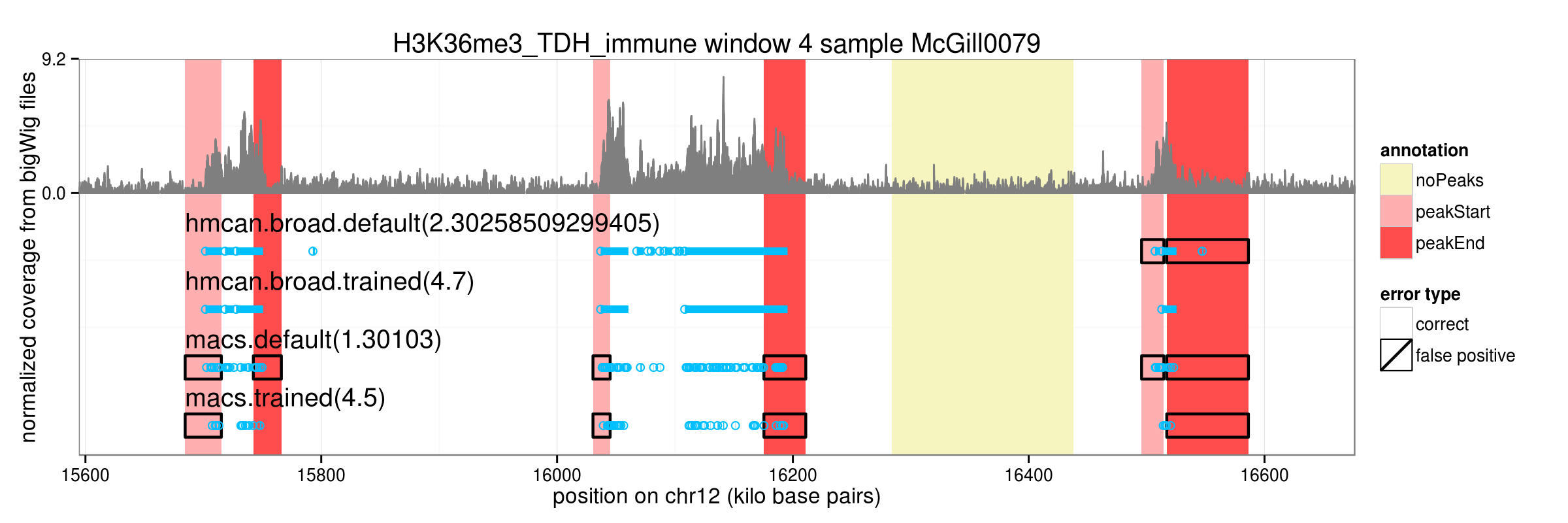}
\end{center}

Shown are default and trained algorithms for samples of 2 different
histone mark types. Consistent with the quantitative results, it is
clear that macs.trained is the best for the H3K4me3 data, and that
hmcan.broad.trained is the best for the H3K36me3 data.
 
\newpage \section*{\figseveral: Windows labeled by 2 different
  annotators}

Annotated regions for 2 different annotators (top=AM, bottom=TDH) on
the same genomic regions and sample sets (H3K36me3 immune cell
types). It is clear that the annotators focus on different levels of
detail, but have in general the same visual definition of a peak.

First, we show just 1 profile:

\begin{center}
  \includegraphics[width=\textwidth]{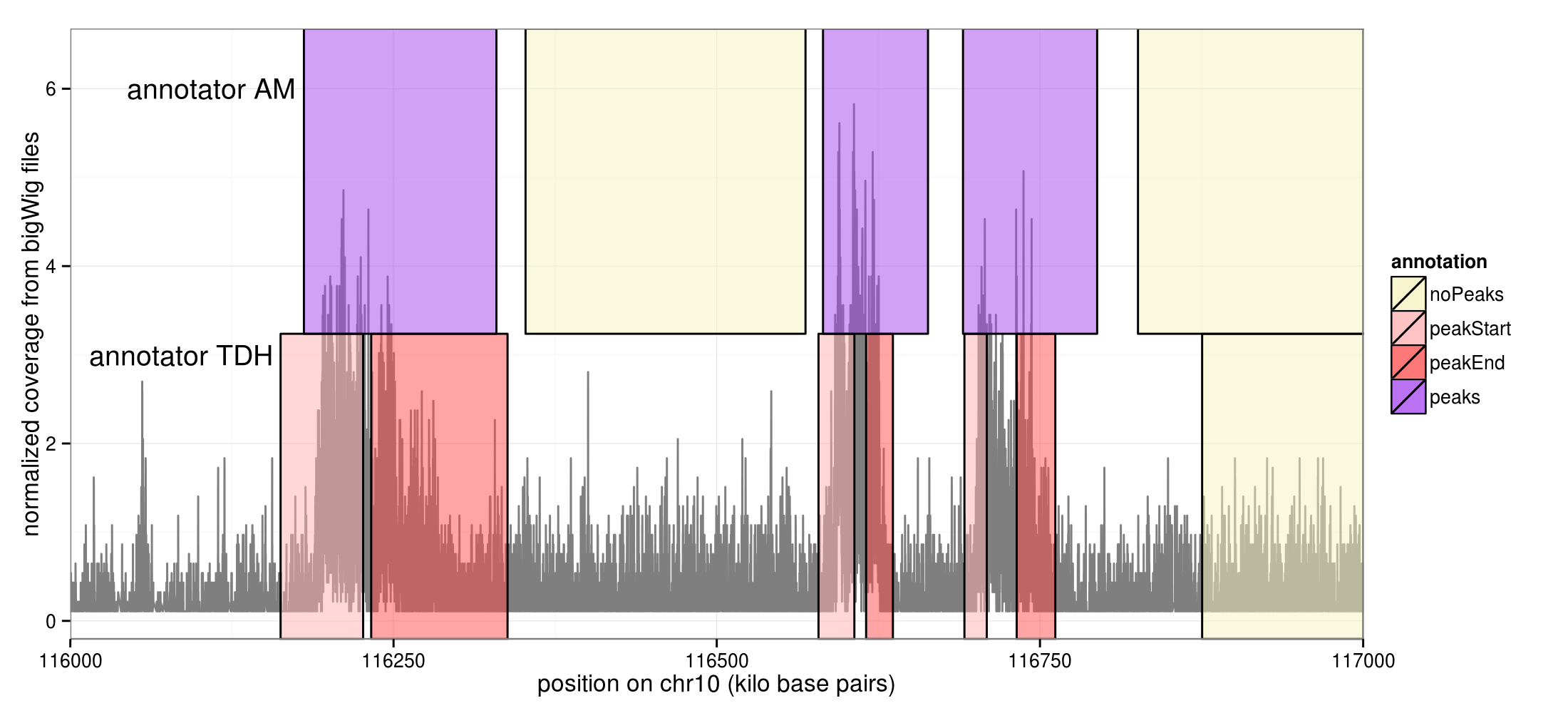}
\end{center}

We also show all profiles, for several annotated windows:

\begin{center}
  \includegraphics[width=\textwidth]{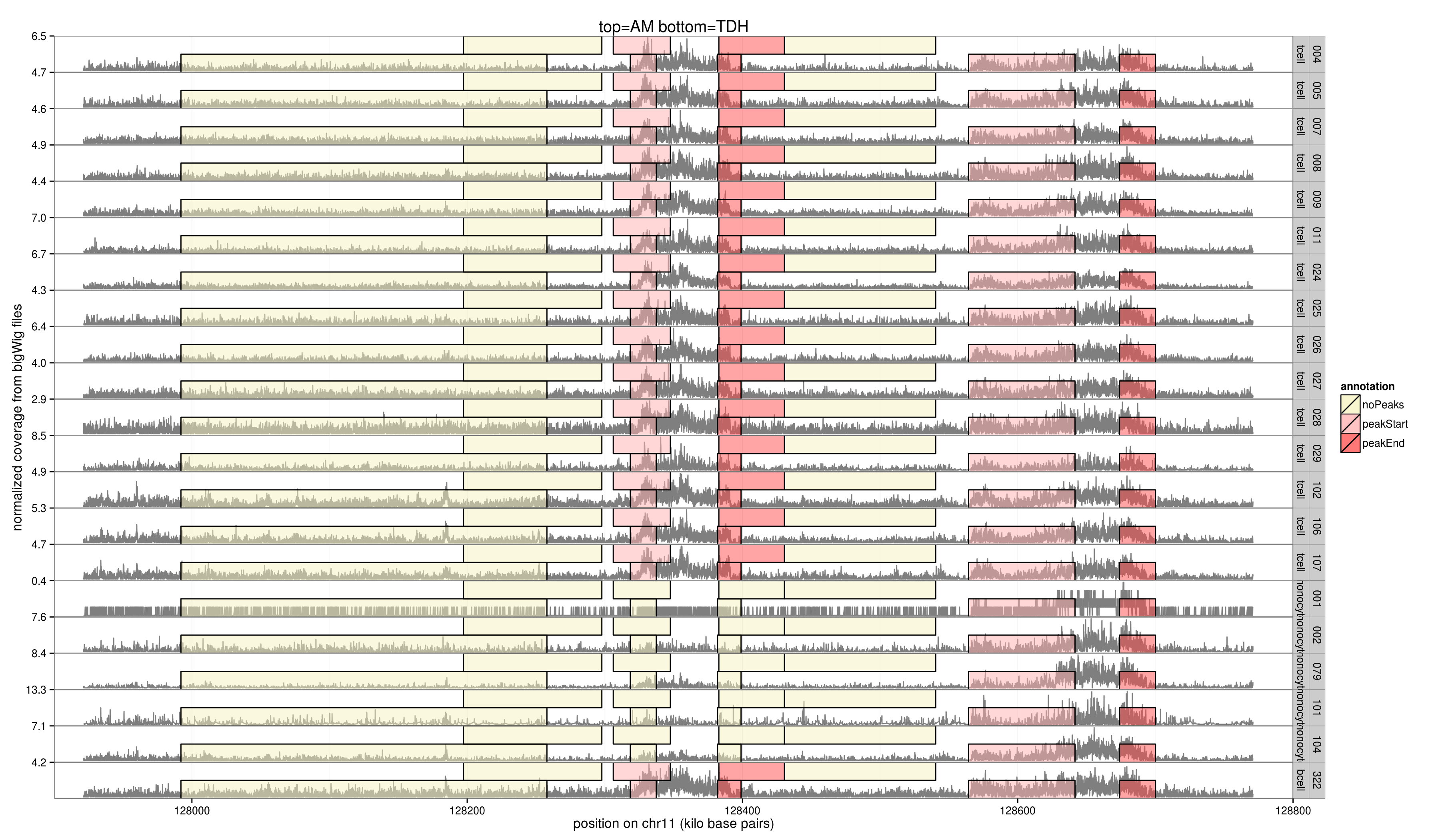}
\end{center}

\begin{center}
  \includegraphics[width=\textwidth]{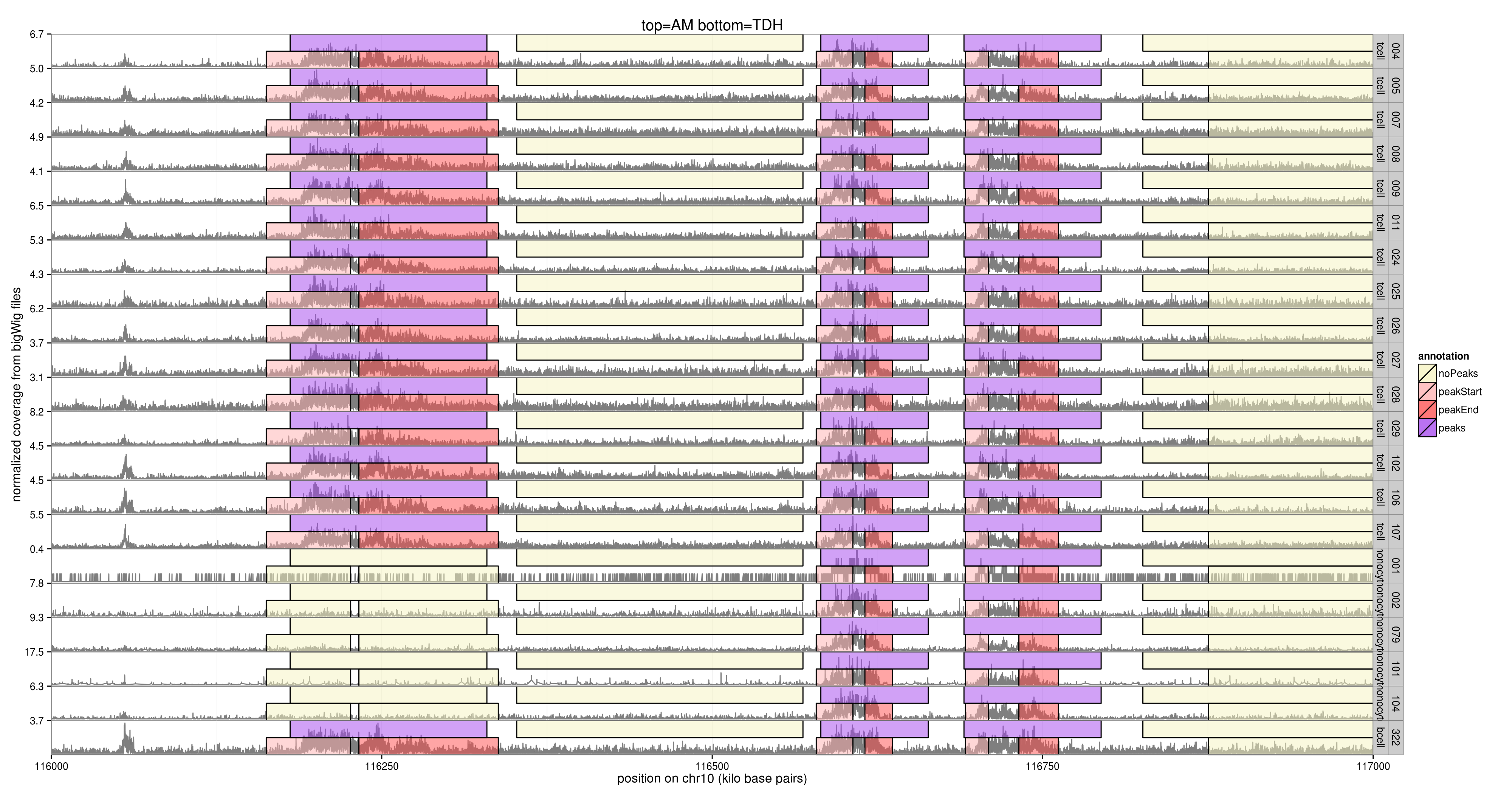}
\end{center}

\begin{center}
  \includegraphics[width=\textwidth]{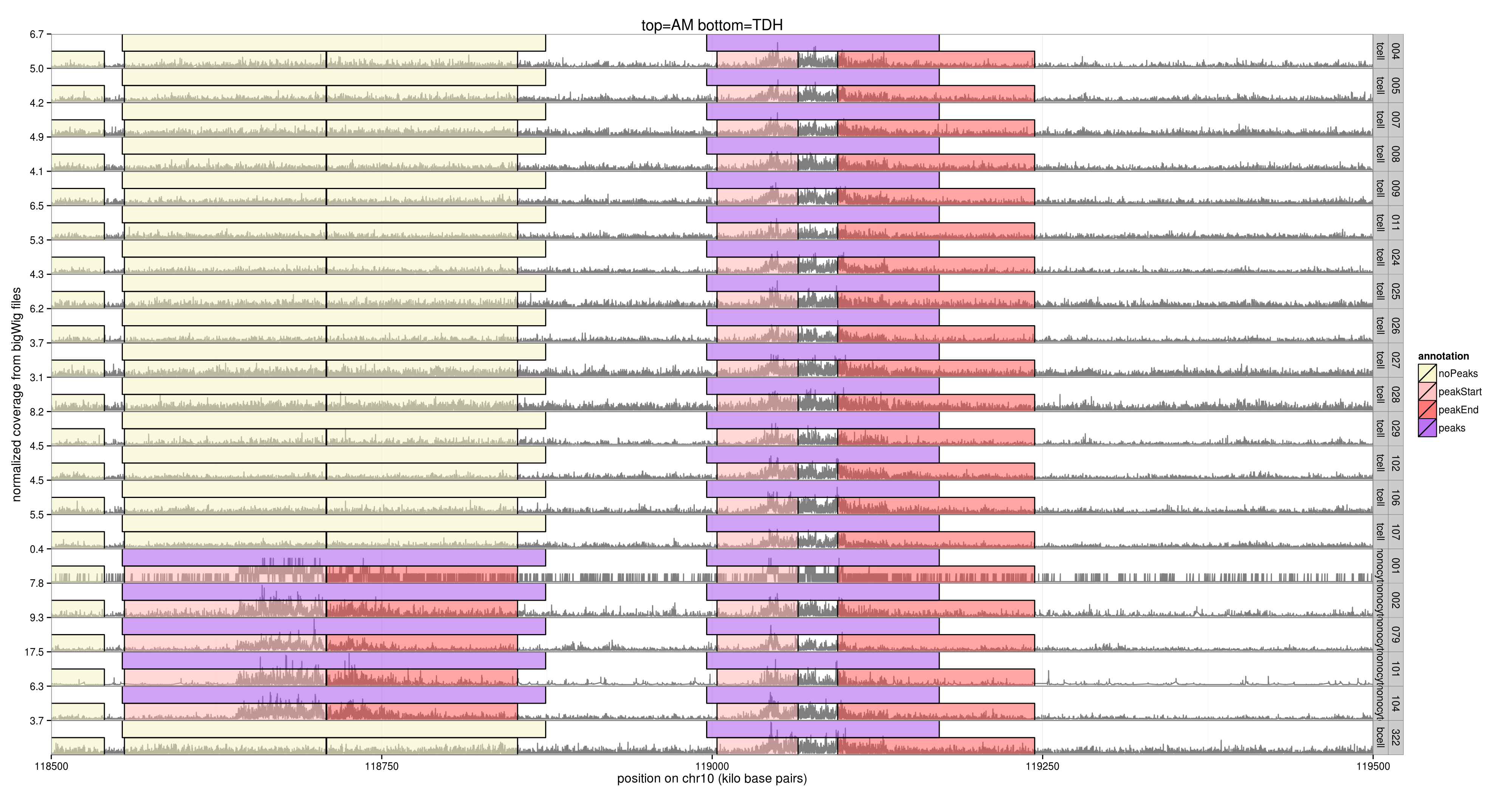}
\end{center}

\begin{center}
  \includegraphics[width=\textwidth]{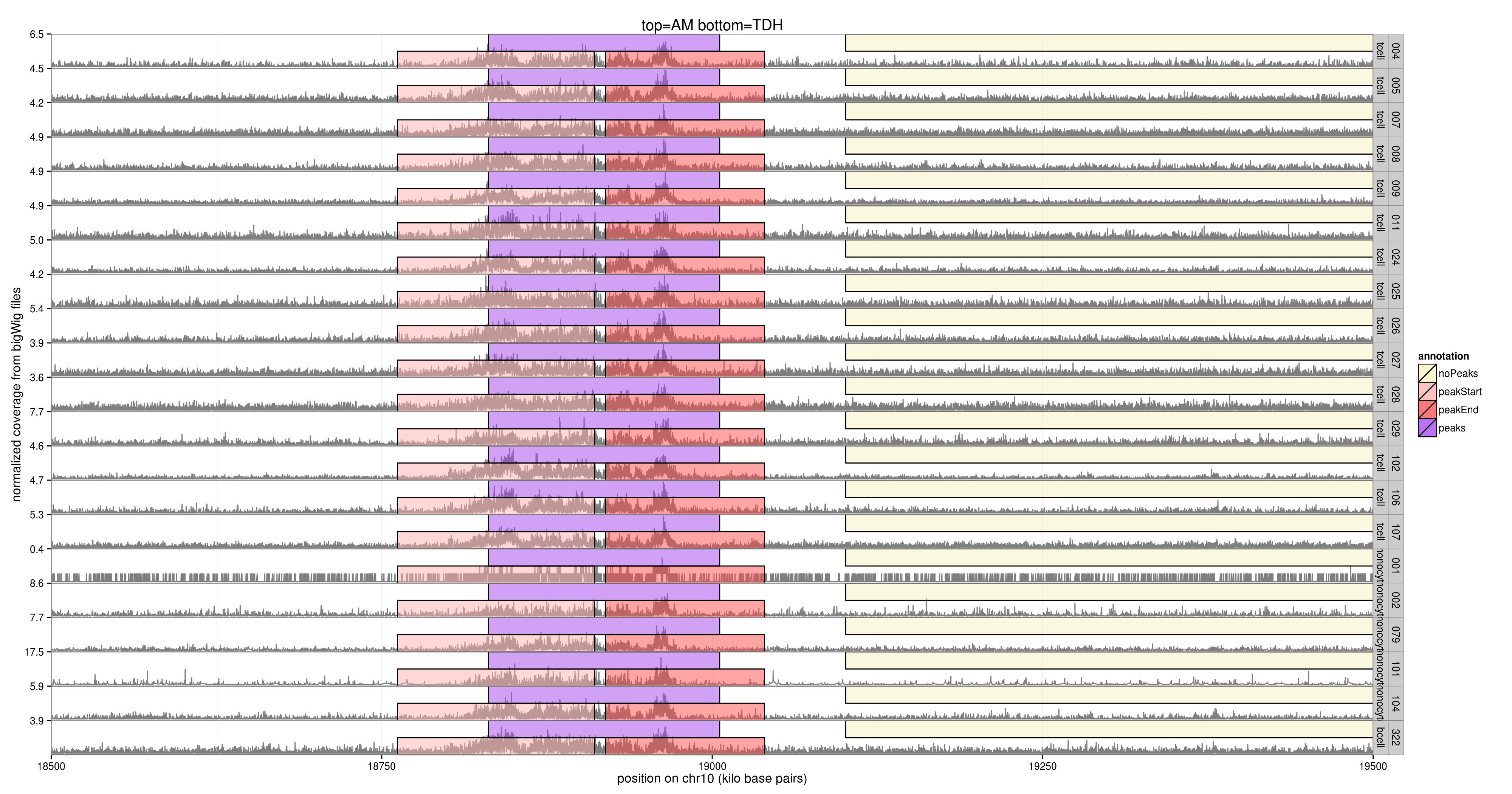}
\end{center}

\begin{center}
  \includegraphics[width=\textwidth]{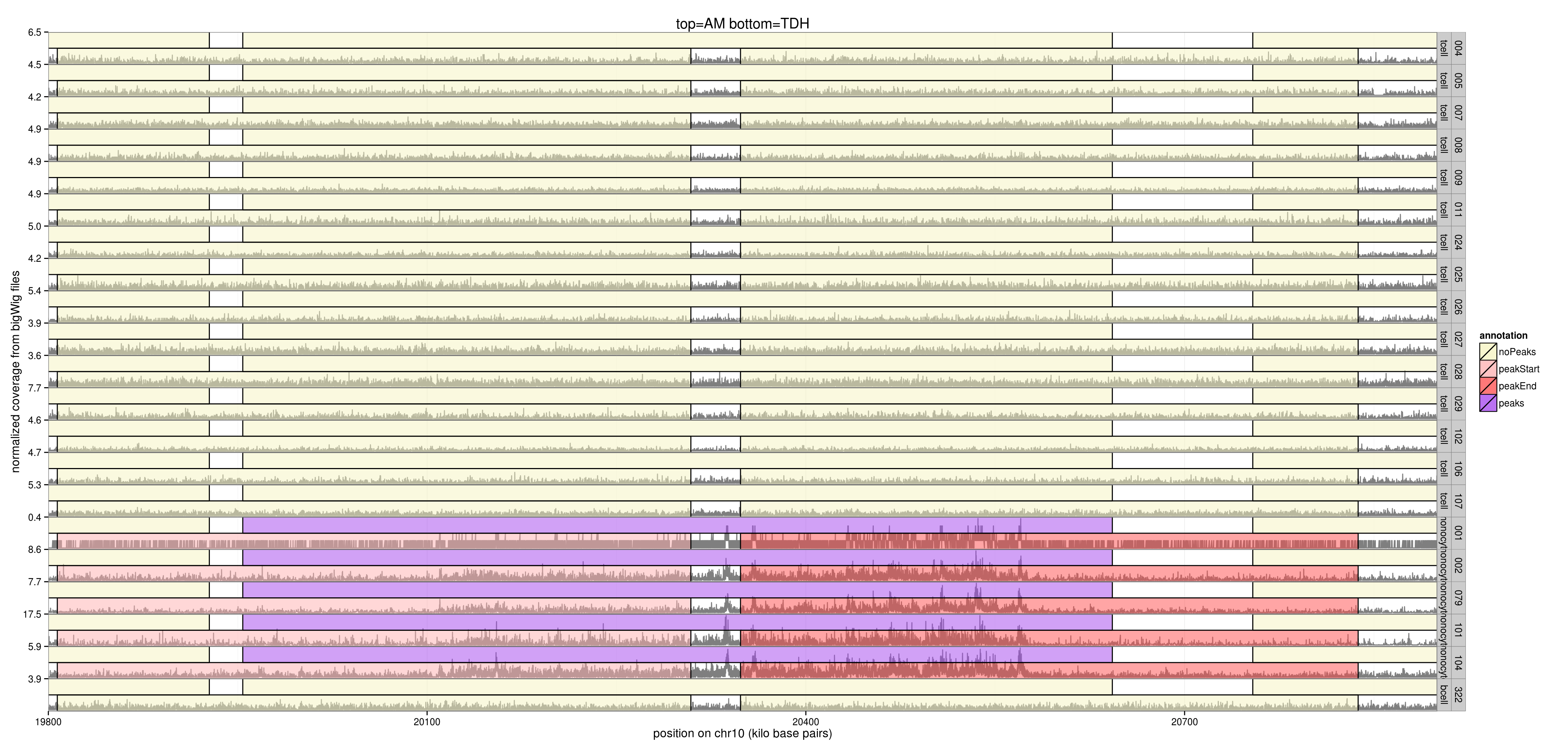}
\end{center}

\begin{center}
  \includegraphics[width=\textwidth]{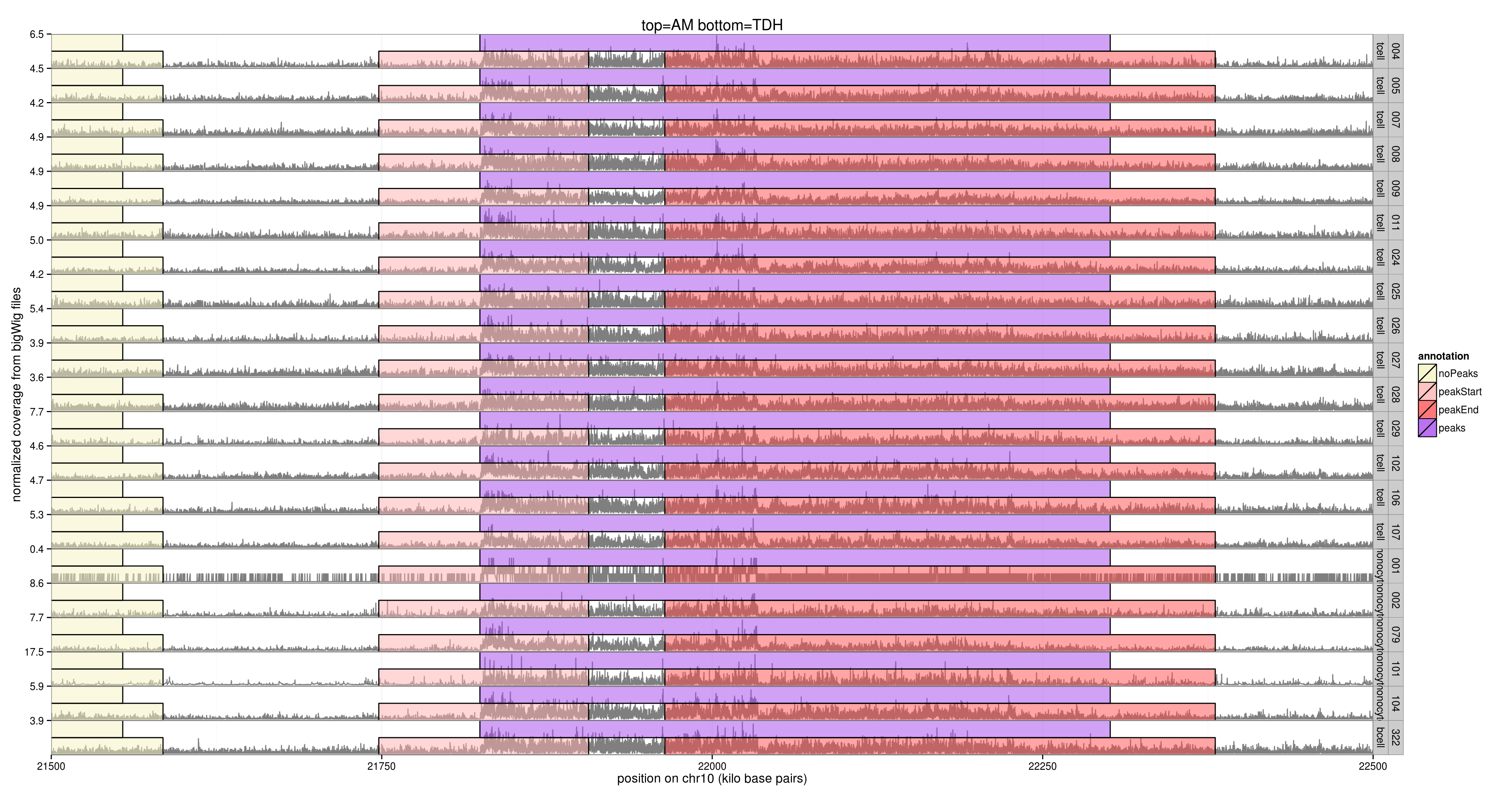}
\end{center}

\begin{center}
  \includegraphics[width=\textwidth]{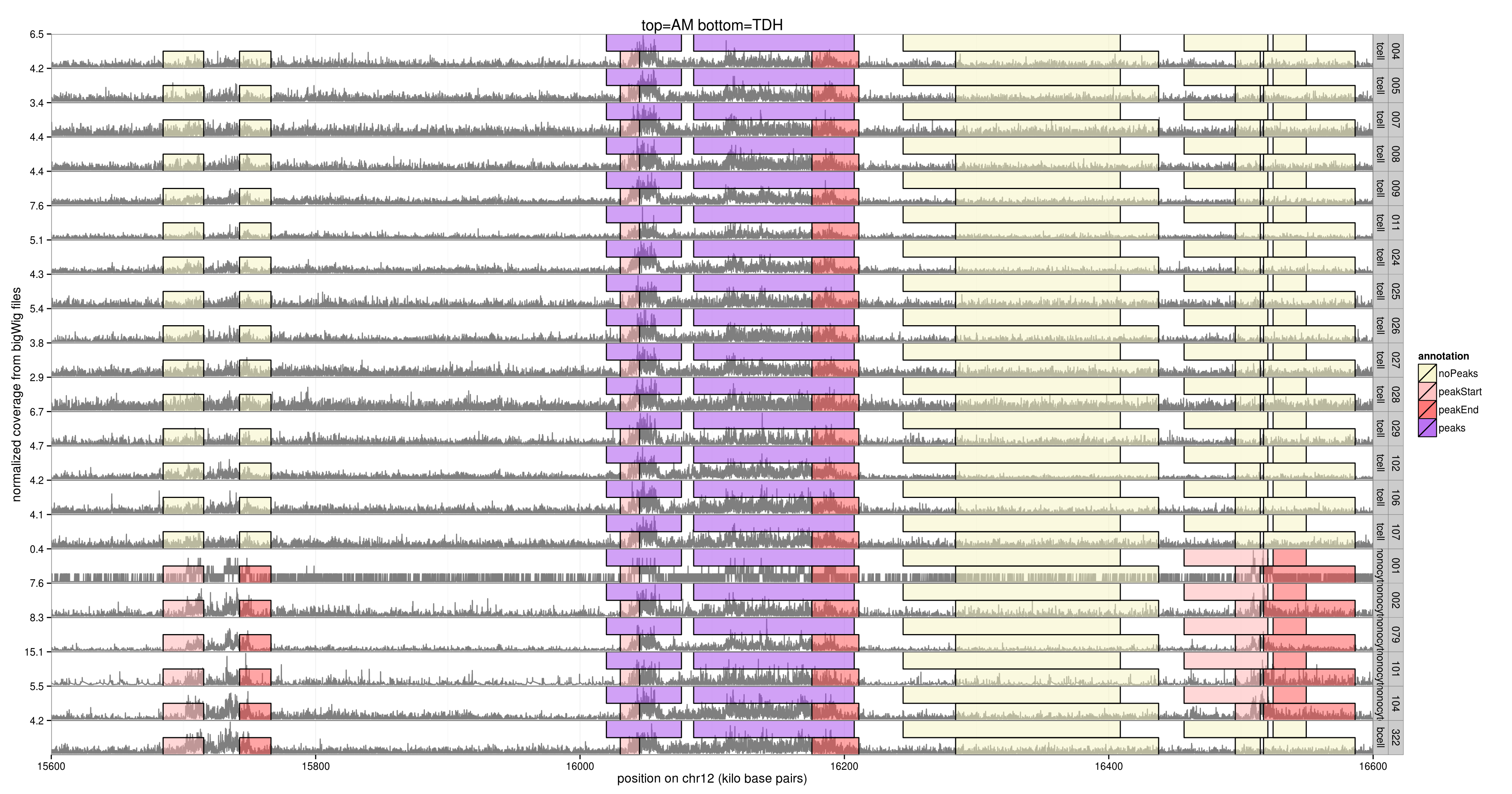}
\end{center}

\newpage \section*{\figannotators: train on one annotator, test on
  another annotator}

\begin{center}
  \includegraphics[width=\textwidth]{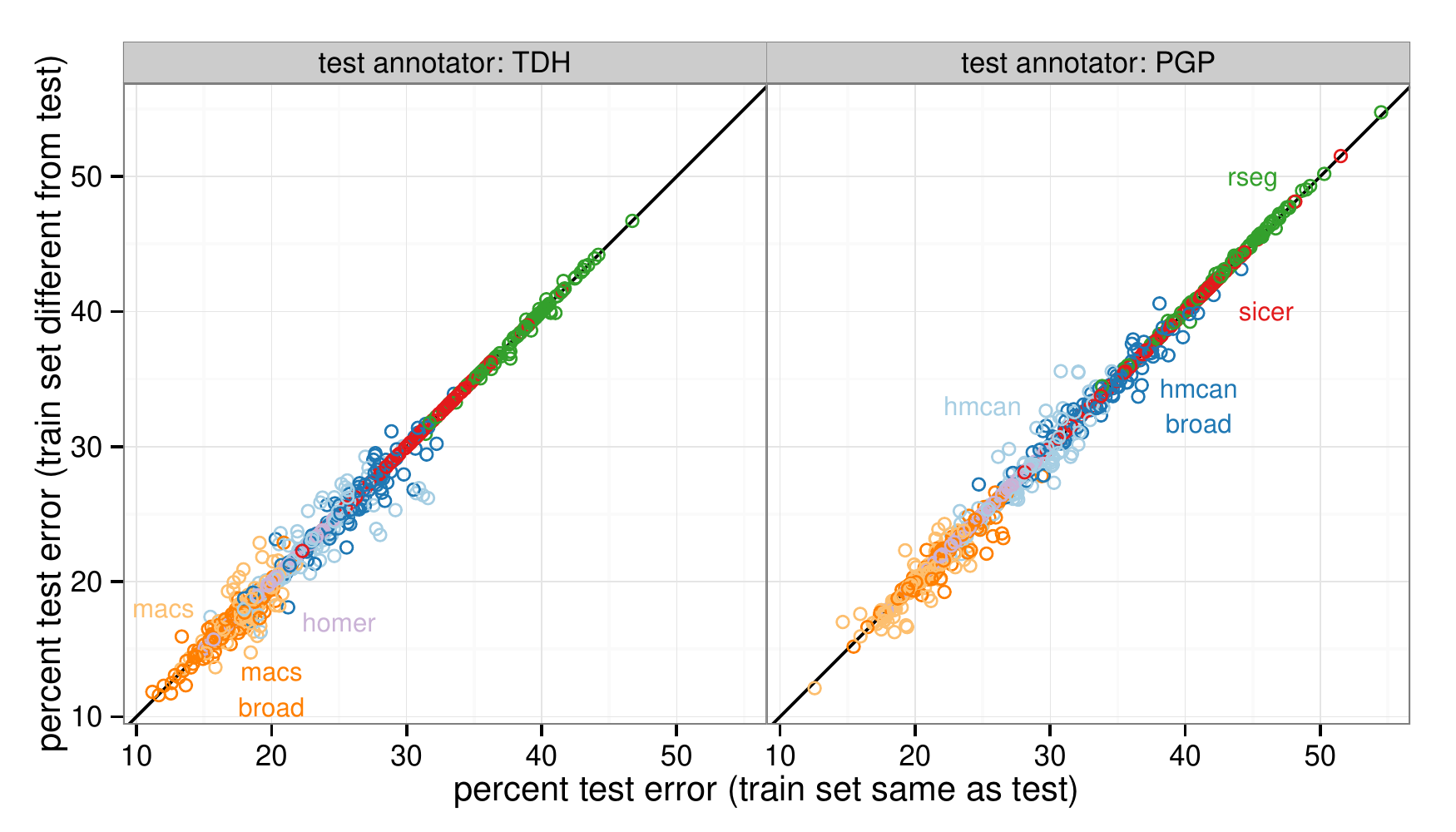}
\end{center}
H3K4me3 immune cell data sets were annotated by TDH and PGP. We
trained models on one or the other annotator, and then tested those
models on the same or a different annotator.
We repeated 100 trials of the following computational experiment: we
randomly designated half of the annotated windows as test data in each
data set, then selected the parameter $\hat \lambda$ with minimum
annotation error on a train set of 10 annotated windows. 

It is clear
for all models that it does not make much difference in terms of test
error when training on one or another annotator.

\newpage
 \section*{\figtypes: train on some cell types, test on
 some other cell types}

\begin{center}
  \includegraphics[width=\textwidth]{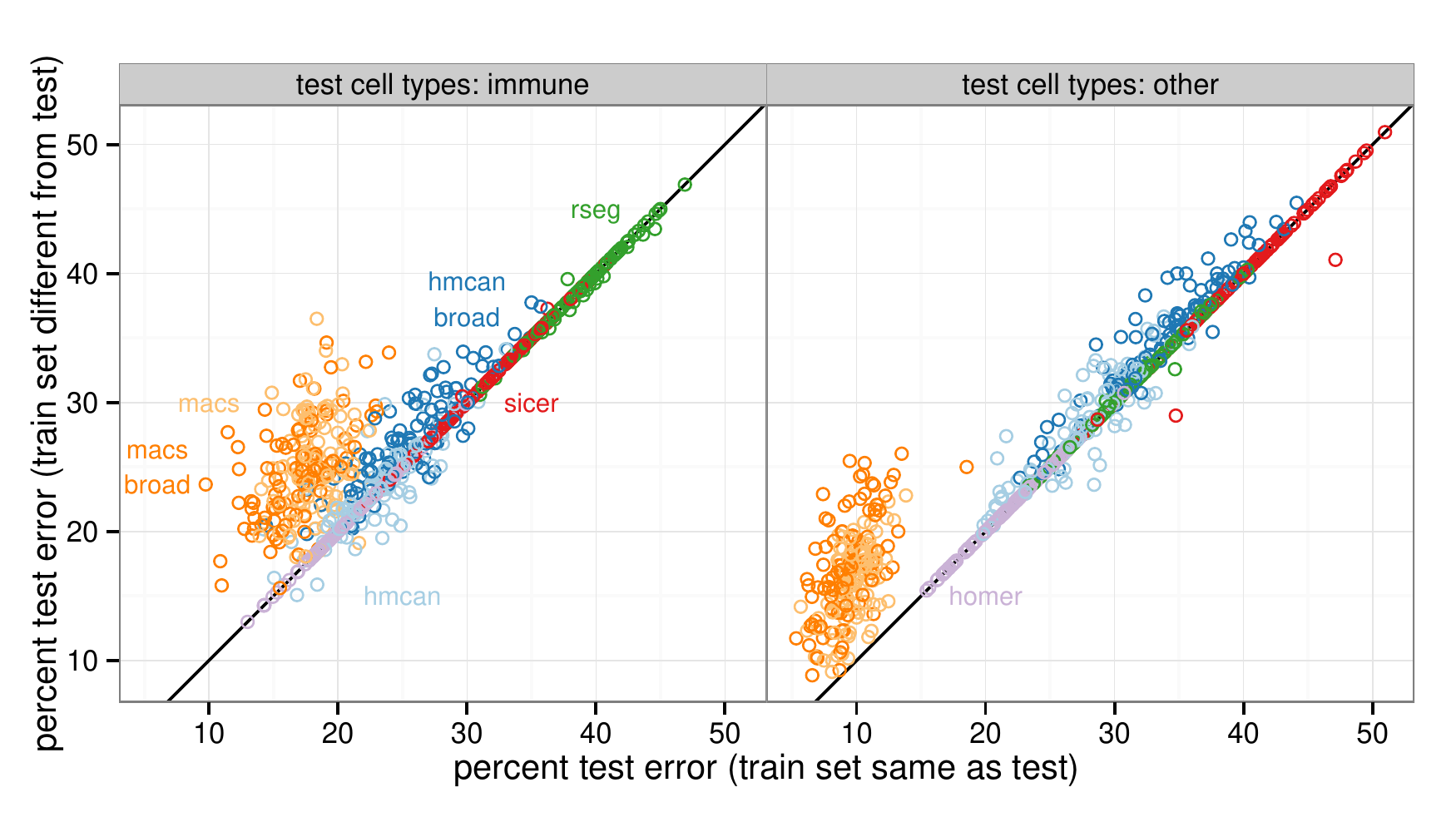}
\end{center}
TDH annotated H3K4me3 data sets of immune and other cell types. We
trained models on one or the other cell types, and then tested those
models on the same or a different set of cell types.  We repeated 100
trials of the following computational experiment: we randomly
designated half of the annotated windows as test data in each data
set, then selected the parameter $\hat \lambda$ with minimum
annotation error on a train set of 10 annotated windows.

It is clear that for some models such as macs and macs.broad, a model
trained on the different cell types (e.g. right panel: train on
immune, test on other) yields higher test error than a model trained
on the same cell types (e.g. train on other, test on other). It is
also clear that the train data do not make much difference for other
models, such as rseg, sicer, hmcan, hmcan.broad, and homer.

\newpage

\bibliographystyle{abbrvnat}
\bibliography{refs}







\end{document}